\documentclass[12pt,preprint]{aastex}

\slugcomment{Revised version from January 15, 2009}

\shorttitle{Circumgalactic metal-line absorbers}
\shortauthors{Richter et al.}

\begin{document}


\title{A population of weak metal-line absorbers \\ 
surrounding the Milky Way
\footnote{Based on observations obtained with the NASA/ESA 
Hubble Space Telescope, which is operated by the Space Telescope Science 
Institute (STScI) for the Association of Universities for Research in Astronomy, 
Inc., under NASA contract NAS5D26555.}}


\author{Philipp Richter\altaffilmark{*}}
\affil{Institut f\"ur Physik und Astronomie, Universit\"at Potsdam,
Haus 28, Karl-Liebknecht-Str.\,24/25, 14476 Potsdam/Golm, Germany}
\email{prichter@astro.physik.uni-potsdam.de}

\author{Jane C. Charlton}
\affil{Department of Astronomy and Astrophysics, Pennsylvania State University,
University Park, PA 16802, USA}

\author{Alessio P. M. Fangano,}
\author{Nadya Ben Bekhti}
\affil{Argelander-Institut f\"ur Astronomie, Universit\"at Bonn,
Auf dem H\"ugel 71, 53121 Bonn, Germany}

\and

\author{Joseph R. Masiero}
\affil{Institute for Astronomy, University of Hawaii, 2680 Woodlawn Drive,
Honolulu, HI 96822, USA}


\altaffiltext{*}{DFG Emmy-Noether-Fellow}


\begin{abstract}

We report on the detection of a population of weak metal-line absorbers
in the halo or nearby intergalactic environment of the Milky Way.
Using high-resolution ultraviolet absorption-line spectra of bright
quasars (QSO) obtained with the {\it Space Telescope Imaging Spectrograph}
(STIS), along six sight lines we have observed unsaturated,
narrow absorption in O\,{\sc i} and Si\,{\sc ii}, together
with mildly saturated C\,{\sc ii} absorption at high radial
velocities ($|v_{\rm LSR}|=100-320$ km\,s$^{-1}$).
The measured O\,{\sc i} column densities lie in the range
$N$(O\,{\sc i}$)<2 \times 10^{14}$ cm$^{-2}$ implying that these
structures represent Lyman-Limit Systems and sub-Lyman-Limit System
with H\,{\sc i} column densities between 
$10^{16}$ and $3 \times 10^{18}$ cm$^{-2}$, thus below
the detection limits of current 21cm all-sky surveys of 
high-velocity clouds (HVCs).
The absorbers apparently are not directly associated with any of the
large high-column density HVC complexes, but
rather represent isolated, partly neutral gas clumps embedded 
in a more tenuous, ionized gaseous medium situated in the halo
or nearby intergalactic environment of the Galaxy.
Photoionization modeling of the observed low-ion ratios suggest
typical hydrogen volume densities of $n_{\rm H}>0.02$ cm$^{-3}$ and
characteristic thicknesses of a several parsec down to sub-parsec scales.
For three absorbers, metallicities are constrained in the range
$0.1-1.0$ solar, implying that these gaseous structures may have 
multiple origins inside and outside the Milky Way. Using supplementary
optical absorption-line data we find for two other absorbers 
Ca\,{\sc ii}/O\,{\sc i} column-density ratios that correspond to
solar Ca/O abundance ratios. This finding indicates that these clouds
do not contain significant amounts of dust.

This population of low-column density gas clumps in the circumgalactic
environment of the Milky Way is indicative of 
the various processes that contribute to
the circulation of neutral gas in the extended halos of spiral galaxies. 
These processes include the accretion of gas from the intergalactic medium 
and satellite galaxies, galactic fountains, and outflows.
We speculate that this absorber population represents the local analog of
weak Mg\,{\sc ii} systems that are commonly observed in the
circumgalactic environment of low- and high-redshift galaxies. 

\end{abstract}



\keywords{ISM: clouds -- quasars: absorption lines -- Galaxy: halo}



\section{Introduction}

As we know from emission and absorption line
measurements, galaxies at low and high redshift
are surrounded by large amounts of gaseous
matter. This circumgalactic gaseous component is believed to
connect the condensed bodies of galaxies with the
filamentary large-scale structure in the Universe
in which they are contained.
The origin and nature of this circumgalactic gas
is manifold.  Among the varied processes that contribute,
as part of the on-going formation and evolution of galaxies,
are outflows from galaxies and gas accretion from the intergalactic
medium (IGM). Studies of the complex interplay between galaxies
and their outer gaseous environment are thus of crucial importance
to understand the evolution of galaxies.

Due to the low densities, multi-phase nature, and clumpiness,
of circumgalactic gaseous structures, emission
measurements (e.g., in H\,{\sc i} 21cm, H$\alpha$)
are limited to relatively nearby galaxies.
Quasar (QSO) absorption-line spectroscopy, in contrast,
allows us to detect both neutral and ionized gas
over eight orders of magnitude in column-density.
Therefore, it is an ideal method to study the outer
gaseous environment of galaxies and their relationship
to the cosmic web at both low {\it and} high redshift. For
absorbers at $z=0$, the most
important diagnostic lines are located in the
ultraviolet (UV) and far-ultraviolet (FUV) and thus
space-based spectrographs such as the Space Telescope
Imaging Spectrograph (STIS) installed on the
{\it Hubble Space Telescope} (HST) or the {\it Far Ultraviolet
Spectroscopic Explorer} (FUSE) are required.
At higher redshifts many of these absorption lines
are shifted into the optical regime and can be observed
with ground-based spectrographs.

The strong Mg\,{\sc ii} resonant doublet near 2800 \AA\, 
has been used extensively to
study strong and weak metal-line absorption in QSO spectra and its
relationship to Galactic structures in the redshift range $0.3<z<2.2$
(e.g., Bergeron \& Boisse 1992; Charlton \& Churchill 1998; Ding et
al.\,2005).  Strong Mg\,{\sc ii} absorbers have rest-frame equivalent
widths $W_{2796} > 0.3$ \AA\, and are commonly found within $35
h^{-1}$ kpc of luminous galaxies.  In contrast, the so-called {\it
weak Mg\,{\sc ii}} systems that have equivalent widths $W_{2796} \leq
0.3$ \AA, appear to be less tightly associated with galaxies and are
typically found at larger impact parameters.  Despite a large
observational and theoretical effort, however, the origin and nature
of the Mg\,{\sc ii} absorbers is not yet fully understood.  It is
quite difficult to determine the relative contributions to the
absorption from various gas-circulation processes such as galactic
winds, tidal galaxy interactions, and accretion of intergalactic gas
from the cosmic web (see, e.g., Rigby et al.\,2002).

Turning to our very local intergalactic gaseous environment,
it has been known for a long time that
the Milky Way is surrounded by a population of neutral/partly
ionized gas clouds with substantial neutral hydrogen column densities
$N$(H\,{\sc i})$>10^{19}$ cm$^{-2}$ (Muller et al.\,1963). These clouds
manifest themselves as high-latitude
gaseous structures seen in H\,{\sc i} 21cm emission
at high radial velocities, $|v_{\rm LSR}| > 100$ km\,s$^{-1}$.
Such high velocities are inconsistent with those expected from a simple
model of Galactic rotation (e.g., Wakker et al.\,1991).
These structures are referred to as "high-velocity clouds" (HVCs)
and they are complemented
by the so-called "intermediate-velocity clouds" (IVCs) which have
somewhat lower radial velocities, in the range
$40 < |v_{\rm LSR}| \leq 100$ km\,s$^{-1}$. Combined 21cm emission
and UV absorption measurements carried out during the last
decade have shown that the metallicities of the IVCs and HVCs range
from $\sim 0.1$ to $\sim 1.0$ times solar 
(Wakker et al.\,1999, 2001; Richter et al.\,1999, 2001; Gibson et al.\,2001;
Tripp et al.\,2003; Collins, Shull \& Giroux 2003).
This abundance range suggests that both the infall of metal-deficient gas and the
circulation of metal-rich gas between disk and halo
contribute to the Milky Way HVCs and IVCs.
It has also been found that the circumgalactic gaseous
environment of the Milky Way is an extreme multiphase medium with gas
temperatures, particle densities, and ionization
fractions spanning several orders of magnitude.

Most of the recent studies on the Milky Way's circumgalactic
gaseous environment have concentrated on the distribution, chemical
composition, and physical properties of the 21cm HVCs/IVCs and
the highly-ionized halo absorbers (e.g., Richter 2006; 
Fox et al.\,2004, 2005; Ganguly et al.\,2005; Tripp et al.\,2003;
Sembach et al.\,2003; Collins, Shull \& Giroux 2006).
Relatively little effort has gone into exploring circumgalactic
neutral and weakly ionized metal absorbers at {\it low} H\,{\sc i}
column densities, i.e., H\,{\sc i} Lyman-Limit Systems (LLS) with
log $N$(H\,{\sc i}$)\leq 19$ and sub-LLS with log $N$(H\,{\sc i}$)
\leq 17.2$ (see also Richter, Sembach \& Howk 2003).
If the gas distribution in the Milky Way halo and in the
Local Group is representative of low-redshift galaxy environments,
one would expect that there exists an extended population of low-column
density, low-ion absorbers around the Milky Way at distances
$30< d <100$ kpc. These would be
the local analogs of weak Mg\,{\sc ii} absorbers, however
this local circumgalactic absorber population has not been
widely studied. These Local Group counterparts of weak
Mg\,{\sc ii} systems are an important element for
placing the Milky Way´s IVC and HVC populations into the more global
context of galaxy/IGM interactions.

Using optical QSO spectra we
recently have detected weak high-velocity Ca\,{\sc ii} and Na\,{\sc i}
absorption that arise from parsec-scale H\,{\sc i} clumps in the Milky
Way halo. These structures have typical H\,{\sc i} column densities in the
range $10^{18}-10^{19}$ cm$^{-2}$ (Richter et al.\,2005;
Ben Bekhti et al.\,2008).
In this paper we continue, at even lower neutral gas column densities, our
analysis of weak metal
absorption in the Milky Way halo and the nearby intergalactic
environment. For this analysis
we have studied unsaturated high-velocity UV absorption from low ions,
such as O\,{\sc i}, Si\,{\sc ii} and C\,{\sc ii}, using STIS high-resolution
spectra of low-redshift QSOs.
As we show here, there indeed exists a population
of circumgalactic LLS and sub-LLS that can be linked to the
Mg\,{\sc ii} absorbers seen in low-and high-redshift QSO spectra.
This paper is organized as follows: in $\S2$  we briefly present the
archival STIS data that is used in this study and describe our
data analysis method.
In $\S3$ we describe the properties of the individual absorption
systems found in the STIS data. The modeling of the ionization
properties of the absorbers and their physical conditions is
presented in $\S4$, and the results in $\S5$.
Finally, a summary is given in $\S6$.

\section{Spectral Data Selection and Analysis Method}

For our study we have made use of archival QSO absorption
line data from HST/STIS, as publically available 
in the MAST\footnote{\tt www.archive.stsci.edu} archive.
More than 30 STIS QSO spectra obtained
with the high-resolution grating E140M are
available in the archive. For some of these sightlines,
additional spectra from the E230M grating are also
available.
The STIS E140M and E230M gratings provide
a spectral resolution of $R\sim 45,000$ and
$\sim 30,000$, respectively. These numbers correspond
to velocity resolutions of $\sim 6.5$ and $\sim 10.0$ km\,s$^{-1}$.
From the archive we have selected 26 QSO E140M spectra 
that have a signal-to-noise ratio (S/N) of $>8$ per 
resolution element in the spectral region near 
the strong O\,{\sc i} line at $1302$ \AA.
Table 1 lists the QSO properties and the HST/STIS grating
information for these sightlines. Fig.\,1 shows
the position of the 26 QSOs plotted on the H\,{\sc i} 
high-velocity sky map from the 21cm Leiden/Argentine/Bonn
(LAB) all-sky survey (Kalberla et al.\,2005).
The STIS data were reduced using the standard STIS
reduction pipeline (Brown et al.\,2002). Separate
exposures were combined following
the procedures described by Narayanan et al.\,(2005).

From a first visual inspection of the metal-line absorption
along the selected QSO sightlines one can immediately 
identify a number of strong  high-velocity absorption features that
coincide in position and radial velocity with high-velocity,
high-column density H\,{\sc i} gas seen in the LAB survey.
Several of these particular
interesting STIS sightlines have been used to probe chemical
abundances in the most prominent large HVC complexes 
(see, e.g., Wakker et al.\,1999; Richter et al.\,2001; Gibson et al.\,2001;
Tripp et al.\,2003; Collins, Shull \& Giroux\,2003). The aim of this paper is,
however, to search for weak, isolated neutral gas absorbers in the halo that are
not directly associated with known 21cm HVC complexes and that have
H\,{\sc i} column densities below the detection limit of typical HVC
21cm observations (i.e., circumgalactic LLS and sub-LLS). It has to be noted that 
is very difficult to directly measure weak high-velocity H\,{\sc i} in the 
Lyman series in the UV and FUV, e.g., using data from the 
{\it Far Ultraviolet Spectroscopic Explorer} (FUSE). One problem is that
for the lower Lyman series lines (Ly\,$\alpha$ to Ly\,$\gamma$) the strong 
local disk absorption completely overlaps any high-velocity H\,{\sc i} absorption 
features. In addition, the presence of molecular hydrogen absorption
in the FUV band $<950$ \AA\, and the (typically) low signal-to-noise ratios
in the FUSE data at these wavelengths strongly hamper the detection of high-velocity
absorption components in the higher Lyman series lines
(however, three counter-examples are discussed later in this paper).

Among low ionization transitions, neutral oxygen (O\,{\sc i}) is
the best indirect tracer
of neutral hydrogen. This is because O\,{\sc i} and H\,{\sc i}
have similar ionization
potentials and in neutral gas they are coupled by a strong
charge-exchange reaction. To identify a possible population 
of isolated, low-column density absorbers in the circumgalactic
environment of the Milky Way 
we therefore have used the strong
O\,{\sc i} transition at $1302.2$ \AA
\footnote{Throughout the paper wavelengths are given in
units of \AA.}
in the E140M waveband together with a number of 
selection criteria that are described in the following. 
First, we have considered only 
halo absorption components at velocities $|v_{\rm LSR}|>
100$ km\,s$^{-1}$ that are detected in O\,{\sc i} $\lambda 1302.2$
absorption. Second, we have considered only those high-velocity O\,{\sc i}
absorbers that have central absorption depths in the
$\lambda 1302.2$ line of less than unity,
corresponding to log $N$(H\,{\sc i})$<18.5$ for gas with an
oxygen abundance of $>0.1$ solar. Third, we have considered only 
those weak, high-velocity O\,{\sc i} absorbers that are not
associated in position and radial velocity with known, large
HVC complexes as seen in the LAB survey (see Fig.\,1).

Using these criteria we have identified seven isolated, weak, high-velocity
O\,{\sc i} absorbers along the six QSO sightlines towards
PG\,1211+143, RX\,J1230.8+0115,
NGC\,4151, Ton\,S210, PHL\,1811 and PG\,1116+215. The positions
of these sightlines are indicated in Fig.\,1. 
Four out of these seven absorbers are located 
near the northern Galactic pole in the range
$l=120-300$ and $b>60$. However, given the overall distribution
of the 26 QSOs in our sample and the positions of high-column density 
HVCs in the sky (Fig.\,1), this apparent clustering is 
not significant statistically.
Table 1 (first section) lists the sightline properties for the six QSOs
where weak high-velocity O\,{\sc i} is detected,
as well as the absorption characteristics of the high-velocity gas
in these directions (last three columns).
In a similar manner, the second section in Table 1 lists the 11 QSO 
sightlines in our sample that have no high-velocity O\,{\sc i} absorption
detected at a high significance level.
The third section of Table 1 lists
all QSO sightlines in our sample that have (mostly strong) high-velocity
absorption in O\,{\sc i} and other ions, 
arising from known, large HVC complexes
seen in 21cm emission (e.g., the Magellanic Stream, Complex C, see 
Fig.\,1). Note that along the line of sight towards NGC\,7469 (last section of Table 1) 
weak O\,{\sc i} absorption is seen at high radial velocities. This 
absorption clearly is caused by the ionized envelope of the Magellanic Stream
in this direction. We do not consider this absorber any further,
as it does not fulfill our third selection criterion of being isolated from
large 21cm HVCs (see above).

We have analyzed the available E140M and E230M STIS data for
PG\,1211+143, RX J1230.8+0115, NGC\,4151, Ton\,S210, and
PHL\,1811 to study the physical properties of the 
high-velocity absorbers in these directions.
The high-velocity absorber towards PG\,1116+215 recently
has been analyzed in detail by Ganguly et al.\,(2005) and
we adopt their results for our discussions in \S5 and \S6.
For the spectral analysis of the other five sightlines mentioned above
we have fitted the local continuum around each 
absorption line using intermediate- and high-order polynomials.
The absorption profiles then were normalized and transformed into a
Local-Standard-of-Rest (LSR) velocity frame. 
Absorption profiles of O\,{\sc i} and other ions for the high-velocity
absorbers towards PG\,1211+143, RX\,J1230.8+0115, NGC\,4151, Ton\,S210, and
PHL\,1811 are shown in Figs.\,2$-$4.
To derive column densities
and Doppler parameters ($b$ values) we have fitted the absorption components
with Voigt profiles using the {\tt FITLYMAN} routine implemented 
in the ESO MIDAS software package (Fontana \& Ballester 1995). 
For the low ions (i.e., all
neutral and singly-ionized species) in the E140M data we have constructed 
for each sightline an absorber model that defines 
the LSR velocity centroids, $v_0$,
and $b$ values of the individual absorption components.
In order to obtain the most accurate results, the best-defined 
absorption lines, from O\,{\sc i} and Si\,{\sc ii}, with the highest
local S/N were selected and fitted simulanteously. After constraining
$v_0$ and $b$ for each absorption component,
all other detected low-ion lines with lower S/N were fitted to this 
model leaving the column density as the only free parameter. The 
measured $b$ values (as listed in Table 2) suggest that the
line widths are dominated by turbulent motions in the gas. 
The contribution from thermal line broadening is small 
for all metal ions. 
Note that the $b$ values derived from individual fits for each low
ion do not differ within the $1\sigma$ error range, implying that
the assumption of a common $b$ value for the low ions is justified and
that these species reside in the same gas phase. Moreover, for the
only neutral species (O\,{\sc i}) the absorption is generally very weak
and located on the linear part of the curve of growth. The derived
O\,{\sc i} column densities therefore are nearly independent of
the choice of $b$ in the measured range $b=3-12$ km\,s$^{-1}$ for
the low ions.
The metal ions and their corresponding UV transitions analyzed in this study 
include C\,{\sc ii} $\lambda 1334.5$, 
C\,{\sc iv} $\lambda\lambda 1548.2,1550.8$,
O\,{\sc i} $\lambda 1302.2$,
Si\,{\sc ii} $\lambda\lambda 1260.4, 1526.7$, Si\,{\sc iii} $\lambda 1206.5$,
Si\,{\sc iv} $\lambda\lambda 1393.8,1402.8$,
Mg\,{\sc ii} $\lambda 2796.4,2803.5$, Fe\,{\sc ii} $\lambda\lambda 2382.8,2600.2$,
and Al\,{\sc ii} $\lambda 1670.8$.
Intermediate and high ions such as Si\,{\sc iii}, Si\,{\sc iv}, and C\,{\sc iv} are
likely to arise in different physical regions than the low ions and
these species consequently were fitted independently of the
low ions. For the high ions, C\,{\sc iv} and Si\,{\sc iv} were coupled  
for the fitting.

All results from our line fitting are summarized in
Table 2. They are discussed in the following section.

A full analysis of the entire STIS data set including
statistical aspects on the frequency and column-density
distribution of all high-velocity metal absorbers, the 
filling factors of the various ions, and various other 
aspects will be presented in a subsequent paper.

\section{Discussion of Individual Absorber Properties}

In this section we discuss the absorption properties of 
the individual high-velocity metal absorbers that are 
detected along the six QSO sight lines.
Along with the UV absorption lines of the above mentioned ions,
we have plotted in Figs.\,$2-4$ the H\,{\sc i} 21cm spectrum 
for each sightline based on data from the LAB survey 
(Kalberla et al.\,2005). As mentioned above, for none of the
high-velocity absorption systems is associated H\,{\sc i} 21cm emission
detected, indicating that the H\,{\sc i} column densities
and/or the angular sizes of these structures are too small
to be detected in current 21cm all-sky surveys. 

\subsection{PG\,1211+143}

Toward PG\,1211+143 ($l=268, b=+74$),
weak high-velocity metal absorption is 
detected in the lines of O\,{\sc i}, C\,{\sc ii}, Si\,{\sc ii}, 
Si\,{\sc iii}, and C\,{\sc iv} at $v_{\rm LSR}=+174$ km\,s$^{-1}$
(see Fig.\,2). The equivalent width of the O\,{\sc i} $\lambda 
1302.2$ absorption in this component is only $\sim 10$ m\AA\, and 
the corresponding (logarithmic) column density is 
log $N$(O\,{\sc i}$)=13.10\pm 0.03$
for a Doppler parameter of $b=7.5$ km\,s$^{-1}$.
Assuming that the gas has an oxygen abundance between $0.1$ and $1.0$
solar, the corresponding H\,{\sc i} column density range 
is log $N$(H\,{\sc i}$)=16.41-17.41$
(throughout the paper we assume a solar oxygen abundance of
(O/H$)_{\sun}=-3.31$; Asplund et al.\,2004). Therefore, this 
absorber probably represents a sub-LLS. C\,{\sc iv} absorption
is very weak ($\sim 26$ m\AA) and Si\,{\sc iv} absorption is not
seen at all. 

A weaker satellite absorption component 
at $v_{\rm LSR}=+191$ km\,s$^{-1}$ is seen in the 
lines of C\,{\sc ii}, Si\,{\sc iii}, and C\,{\sc iv}. The fact
that, besides C\,{\sc ii}, only intermediate and high ions 
(Si\,{\sc iii} and C\,{\sc iv}) are seen suggests that this 
satellite component is predominantly ionized. There is another
possible absorption system near $+270$ km\,s$^{-1}$, seen in
Si\,{\sc iii}, but it cannot be verified by C\,{\sc ii} since
absorption by C\,{\sc ii}$^{\star}$ in the Galactic disk
dominates at that wavelength. Thus, the status of the possible absorber
at $+270$ km\,s$^{-1}$ in the Si\,{\sc iii} restframe remains
uncertain.

Using FUSE data, Sembach et al.\,(2003) investigated
high-velocity O\,{\sc vi} absorption in the direction of 
PG\,1211+143, but did not find any halo O\,{\sc vi} above
a column-density limit of log $N$(O\,{\sc vi}$)=13.61$. There
is a compact high-velocity cloud (CHVC\,$262.6+75.1+100$) 
only 1.5 degree away from PG\,1211+143 with a radial velocity 
of $+100$ km\,s$^{-1}$ (de Heij et al.\,2002). Given the 
observed velocities, this CHVC may be 
be physically related to the sub-LLS toward PG\,1211+143.
However, deeper H\,{\sc i} observations would be required to
confirm this hypothesis.

\subsection{RX\,J1230.8+0115}

In the direction of RX\,J1230.8+0115 ($l=291, b=+64$) high-velocity
metal absorption in multiple ions is present in two independent components 
at $v_{\rm LSR}=+293$ km\,s$^{-1}$
and $v_{\rm LSR}=+110$ km\,s$^{-1}$ (Fig.\,2).
The absorber at $v_{\rm LSR}=+293$ km\,s$^{-1}$ is seen in
the lines of O\,{\sc i}, C\,{\sc ii}, Si\,{\sc ii},
Si\,{\sc iii}, and C\,{\sc iv}. For the low ions, the best fit
is obtained with a $b$ value of $4.0$ km\,s$^{-1}$, while for 
the intermediate and high ions (Si\,{\sc iii} and C\,{\sc iv}) we
obtain $b=10.0$ km\,s$^{-1}$ as the best solution. The measured
O\,{\sc i} column density (log $N$(O\,{\sc i}$)=13.80\pm 0.03$) corresponds
to an H\,{\sc i} column density range of log $N$(H\,{\sc i}$)=17.11-18.11$,
assuming a $0.1-1.0$ solar oxygen abundance. Given these numbers, this 
system most likely is a LLS.
The radial velocity ($\sim 300$ km\,s$^{-1}$) of this absorber is relatively
high for Galactic halo gas. However, since there is no
bright galaxy close to this sight line, an association with the
Milky Way (or at least with the Local Group) is most likely.
Note that the two additional absorption components at $+260$ and $+210$ 
km\,s$^{-1}$ in the C\,{\sc ii} restframe again belong to 
C\,{\sc ii}$^{\star}$ in the local disk gas. 

The second stronger halo metal system along this sight line
at $v_{\rm LSR}=+110$ km\,s$^{-1}$ is detected in 
C\,{\sc ii}, Si\,{\sc ii}, and Si\,{\sc iii} absorption. 
Unfortunately, at this velocity O\,{\sc i} $\lambda 1302.2$ is blended
with an intervening absorber, but a relatively
stringent $3\sigma$ upper limit of log $N$(O\,{\sc i}$)<13.65$ can
be placed using a $b$ value of $8.0$ km\,s$^{-1}$, as derived
from the fit of the above mentioned low ions.
The Si\,{\sc ii} $\lambda 1260.4$ absorption looks
extremely narrow (only 2 pixels broad); this 
must be caused by the intrinsic noise in the spectrum. 

No high-velocity O\,{\sc vi} data has been published for the 
sightline of RX\,J1230.8+0115, but weak halo O\,{\sc vi} absorption 
is found at a total column density of log $N=13.68$
in the velocity range between $+105$ and $+260$ 
km\,s$^{-1}$ towards the bright QSO 3C\,273.0,
only about one degree away (Sembach et al.\,2003). 
The O\,{\sc vi} measurements of 3C\,273.0 thus
indicate the presence of highly-ionized halo gas in this 
general direction with a similar radial velocity
and suggest a possible physical connection
between high-velocity O\,{\sc vi} and the various
high-velocity absorption components toward RX\,J1230.8+0115.
No low-ion absorption is seen at high velocities in the
spectrum of 3C\,273.0 (see Table 1).
The presence of high-velocity O\,{\sc i} and Si\,{\sc ii} absorption
in the spectrum of RX\,J1230.8+0115 and the absence of similar
absorption features along the 3C\,273.0 sightlines set 
a limit for the transverse size, $l_{\rm t}$, of the 
O\,{\sc i} absorbing structures. 
For an assumed absorber distance of $d=10$ kpc, the non-detection
implies a transverse absorber size of $l_{\rm t}<160$ pc, while for
$d=50$ kpc we obtain $l_{\rm t}<800$ pc. As we will see later,
these size limits are fully consistent with size estimates
from the ionization modeling of the O\,{\sc i} absorbers (\S4.2). 

There is no known CHVC at positive velocities, within 2 degrees 
of the position of RX\,J1230.8+0115, that could be associated 
with this LLS (de Heij et al.\,2002).

\subsection{NGC\,4151}   

The line of sight towards NGC\,4151 ($l=155, b=+75$) has
one high-velocity metal absorber at
$v_{\rm LSR}=+145$ km\,s$^{-1}$ (Fig.\,3). 
There are both 140M and 230M STIS echelle
data available for this sight line. Detected ions include
O\,{\sc i}, C\,{\sc ii}, Si\,{\sc ii},
Si\,{\sc iii}, C\,{\sc iv}, and Si\,{\sc iv} (E140M data), as well as
Mg\,{\sc ii} and Fe\,{\sc ii} (E230M data). The good S/N in the spectrum
suggests that this system is a well-defined
single-component absorber with no evidence for velocity sub-structure.
The $b$ value that we derive from the simultaneous fit of the low
ions is only $3.0$ km\,s$^{-1}$ (and could be lower if the line
is unresolved), while for the intermediate and 
high ions we find $b=11.0$ km\,s$^{-1}$ (Si\,{\sc iii}) and
$b=16.3$ km\,s$^{-1}$ (C\,{\sc iv}, and Si\,{\sc iv}), respectively.
These vastly different Doppler parameters possibly suggest
a pronounced core-envelope structure with a relatively compact neutral
core embedded in a more extended ionized gas layer. The 
O\,{\sc i} column density is log $N$(O\,{\sc i}$)=12.90\pm 0.03$, 
which translates to an H\,{\sc i} column density range of 
log $N$(H\,{\sc i}$)=16.21-17.21$ for
a $0.1-1.0$ solar oxygen abundance. This system therefore is 
a sub-LLS. This absorber is one of two in our sample that 
exhibits narrow, weak Mg\,{\sc ii}
absorption in a single absorption component. We measure an equivalent
width of $W_{\lambda 2796}=48\pm5$ m\AA\, for the stronger of the
two Mg\,{\sc ii} lines at $2796.4$ \AA. This system therefore
mimics the absorption characteristics of weak Mg\,{\sc ii} absorbers
that are frequently found in QSO spectra at low and high redshift
(e.g., Milutinovi\a'c et al.\,2006). There is neither high-velocity
O\,{\sc vi} absorption detected towards NGC\,4151 
(log $N$(O\,{\sc vi}$)\leq 13.42$; Sembach et al.\,2003), nor is there
any CHVC known to exist within 2 degrees of the sightline (de Heij et al.\,2002). 

\subsection{Ton\,S210} 

Toward Ton\,S210 ($l=225, b=-83$) we observe moderately strong metal 
absorption at $v_{\rm LSR}=-167$ km\,s$^{-1}$ 
in the lines O\,{\sc i}, C\,{\sc ii}, 
Si\,{\sc ii}, Mg\,{\sc ii}, Fe\,{\sc ii}, and
Si\,{\sc iii} (Fig.\,4). As with NGC\,4151, E140M and E230M data 
are available for the Ton\,S210 sightline. No high ions are detected in this system,
but the S/N is relatively low near $1550$ \AA\, 
($\sim 4$ per pixel), so that weak
C\,{\sc iv} absorption may be hidden in the noise.
We find $b=6.3$ km\,s$^{-1}$ and log $N$(O\,{\sc i}$)=14.02\pm 0.04$,
implying log $N$(H\,{\sc i}$)=17.22-18.33$ for a $0.1-1.0$ solar
oxygen abundance; this absorber therefore represents a LLS.
C\,{\sc ii} absorption is highly saturated and thus the derived column density
is relatively uncertain (see Table 2). A similar problem occurs for 
Si\,{\sc iii}, which is saturated and has low S/N.
We refrain from deriving a Si\,{\sc iii} column density because of this large uncertainty.
The C\,{\sc ii} profile is asymmetric, implying that there are at
least two weak satellite components located near $-190$ and $-210$ km\,s$^{-1}$ 
(see Fig.\,4). Also the Mg\,{\sc ii} $\lambda 2796.4$ absorption indicates
the presence of a second component near $-190$ km\,s$^{-1}$, 
but this feature most likely is an artifact caused by the noise in the data.
The total equivalent width of the high-velocity Mg\,{\sc ii} $\lambda 2796.4$ absorption
in the range $-215$ to $-130$ km\,s$^{-1}$ is measured to be $303\pm 44$ m\AA. With 
this equivalent width, the system lies exactly on the transition point at 
$0.3$ \AA\, separating weak and strong Mg\,{\sc ii} systems.

The high-velocity absorber towards Ton\,S210 is known to be related to 
the compact high-velocity cloud CHVC\,$224.0-83.4-197$, as seen in
high-resolution 21cm Parkes HIPASS data (Putman et al.\,2002).
Sembach et al.\,(2002) previously have analyzed 
absorption by O\,{\sc i}, C\,{\sc ii}, and 
C\,{\sc iii} in this halo cloud using FUSE spectra of Ton\,S210.
From analysis of the O\,{\sc i} $\lambda 1039.2$ line (which unfortunately
is blended with Galactic H$_2$ absorption) they derive a conservative 
upper limit for the O\,{\sc i} column density in the high-velocity gas of 
log $N$(O\,{\sc i}$)<14.74$.
The O\,{\sc i} column density limit from the FUSE data
is thus $\sim 0.7$ dex above the accurate value of $N$(O\,{\sc i}) 
that we derive from the unblendend, unsaturated $\lambda 1302.2$ 
absorption in the STIS data.
Using the Parkes 21cm data and constraints from the H\,{\sc i} Lyman series 
in the FUSE band pass Sembach et al.\, derive an H\,{\sc i} column density of 
log $N$(H\,{\sc i}$)=18.34\pm0.04$ for this HVC. From the measured O\,{\sc i}/H\,{\sc i}
ratio it is possible to obtain a new estimate of the metallicity of 
this cloud, as will be discussed in detail in $\S4.2$.

There is another high-velocity metal-absorber located near $-242$ km\,s$^{-1}$.
This system must be highly ionized,
as the absorption pattern is dominated by Si\,{\sc iii}, 
C\,{\sc iv} and Si\,{\sc iv}, while among the low ions, only 
C\,{\sc ii} is detected. Other weak highly-ionized absorption components
possibly are present in the range $-150$ to $-50$ km\,s$^{-1}$ (see Fig.\,4). 
This complex intermediate- and high velocity absorption pattern may partly be related
to highly ionized Galactic-fountain material in the disk-halo interface;
however, they are not considered any further in the present analysis.
Sembach et al.\,(2003) find high-velocity O\,{\sc vi} absorption with FUSE towards
Ton\,S210 in the velocity range between $-250$ and $-120$ km\,s$^{-1}$ 
with a total column density of log $N$(O\,{\sc vi}$)=13.97$. The high-velocity
O\,{\sc vi} absorption most likely is related to the various C\,{\sc iv} and 
Si\,{\sc iv} absorption components seen in the STIS data and may indicate 
the highly-ionized outer boundary of the CHVC/LLS towards Ton\,S210
(see also Sembach et al.\,2003 for a detailed discussion of this
sightline).

\subsection{PHL\,1811}     

The line of sight toward PHL\,1811 ($l=47,b=-45$) consists of at least four
high-velocity metal absorbers at negative velocities (Fig.\,4). 
The strongest neutral 
gas absorber is located at $-207$ km\,s$^{-1}$ and is detected in 
O\,{\sc i}, C\,{\sc ii}, Si\,{\sc ii}, Al\,{\sc ii},
Si\,{\sc iii}, C\,{\sc iv}, and Si\,{\sc iv}. For the low ions
we find a single-component Doppler parameter of $b=9.2$ km\,s$^{-1}$. Absorption in the
intermediate and high ions is substantially saturated, so that the
column densities remain unknown or relatively uncertain 
(see Table 2). We measure an O\,{\sc i} column density of
log $N$(O\,{\sc i}$)=14.11\pm 0.05$, which corresponds
to an H\,{\sc i} column density range of log $N$(H\,{\sc i}$)=17.42-18.42$
for a $0.1-1.0$ solar oxygen abundance. This system therefore is a LLS.
Weak, high-velocity O\,{\sc i} absorption
is also seen at $-163$ km\,s$^{-1}$, together with absorption by 
C\,{\sc ii}, Si\,{\sc ii}, Al\,{\sc ii}, Si\,{\sc iii}, C\,{\sc iv}, and Si\,{\sc iv}.
For this component we obtain log $N$(O\,{\sc i}$)=13.23\pm 0.07$
and $b=9.6$ km\,s$^{-1}$ (log $N$(H\,{\sc i}$)=16.54-17.54$; sub-LLS). In addition,
high-velocity absorption is detected in C\,{\sc ii} and Si\,{\sc iii} at
$-264$ km\,s$^{-1}$ and in Si\,{\sc iii}, C\,{\sc iv}, and Si\,{\sc iv}
near $-350$ km\,s$^{-1}$ (see Table 2 for details). 

FUSE observations of PHL\,1811 indicate the presence of considerable amounts
of high-velocity O\,{\sc vi} in the ranges $-360$ to $-295$ km\,s$^{-1}$
(log $N$(O\,{\sc vi}$)=13.85$) and $-200$ to $-65$ km\,s$^{-1}$
(log $N$(O\,{\sc vi}$)=14.38$; Sembach et al.\,2003). 
These velocity ranges suggest a physical
connection between the high-velocity O\,{\sc vi} absorption and the two strong 
C\,{\sc iv} absorption components near $-350$ and $-163$ km\,s$^{-1}$. 
Sembach et al.\,favor a Local Group origin for these highly-ionized 
absorbers towards PHL\,1811.
As already pointed out by Jenkins et al.\,(2003) in their analysis of
the PHL\,1811 sight line (concentrating mostly on absorption at higher redshifts), 
the various low- and high-ion absorption components at high velocities
may also be related to the 21cm HVC
complex GCN. Complex GCN is located in the general direction of PHL\,1811 
and has radial velocities in the range $-300$ to $-200$ km\,s$^{-1}$
(see Wakker 2001 for a detailed description). Also the Parkes HIPASS
21cm survey (Putman et al.\,2002) shows the presence of high-velocity, 
low-column density H\,{\sc i} gas near $-275$ km\,s$^{-1}$ in this 
direction only $\sim 1.2$ degree away (HVC\,$046.1-45.4-275$). It seems
likely that the high-velocity H\,{\sc i} gas seen with
HIPASS is related to the observed 
complex high-velocity absorption pattern between $-200$ and $-300$
km\,s$^{-1}$ towards PHL\,1811 and is part of a larger gaseous complex.

High-velocity absorption towards PHL\,1811
has also been studied in the optical lines of Ca\,{\sc ii}
$\lambda\lambda 3934.8,3969.6$ and Na\,{\sc i}
$\lambda\lambda 5891.6,5897.6$ using high-resolution 
echelle spectra from VLT/UVES (Ben\,Bekhti et al.\,2008)
\footnote{Note that Ben\,Bekhti et al.\,use the name 
J2155$-$0922 instead of PHL\,1811.}.
It is interesting that both high-velocity O\,{\sc i} components,
at $-207$ and $-163$ km\,s$^{-1}$,
are detected in Ca\,{\sc ii} (but not in Na\,{\sc i}).
In Fig.\,5 we show the velocity profiles of the two Ca\,{\sc ii}
lines towards PHL\,1811, based on the VLT/UVES spectra. From 
line-profile fits of the high-velocity Ca\,{\sc ii} absorption
we obtain column densities of 
log $N$(Ca\,{\sc ii}$)=11.75\pm 0.02$ for the cloud at
$-207$ km\,s$^{-1}$ and log $N$(Ca\,{\sc ii}$)=11.13\pm 0.06$
for the one at $-163$ km\,s$^{-1}$.
The simultaneous detection of O\,{\sc i} and Ca\,{\sc ii} in the
two high-velocity absorbers towards PHL\,1811 supports the idea that
the analysis of high-velocity Ca\,{\sc ii} absorption in optical spectra
is a powerful method to investigate the distribution of neutral
gas in the Milky Way halo at H\,{\sc i} column densities above and 
below the detection limit of current 21cm all-sky surveys
(Ben\,Bekhti et al.\,2008). From the measured Ca\,{\sc ii}/O\,{\sc i}
column density ratio it is also possible to obtain information
about the dust content of the high-velocity absorbers towards
PHL\,1811 (see \S4.4).

\subsection{PG\,1116+215}

Towards PG\,1116+215 ($l=223, b=+68$) high-velocity
metal absorption is detected in the range
$v_{\rm LSR}=180-195$ km\,s$^{-1}$
in various low and high ions (O\,{\sc i}, C\,{\sc ii}, Si\,{\sc ii},
Mg\,{\sc ii}, Fe\,{\sc ii}, Si\,{\sc iii}, and C\,{\sc iv}).
As mentioned previously, we here adopt the results from 
Ganguly et al.\,(2005), who present a very detailed study of 
the halo absoption in this direction using the available STIS E140M and E230M data. 
For the main low-ion absorption component at $+193$ km\,s$^{-1}$
Ganguly et al.\,(2005) derive an O\,{\sc i} column density of
log $N$(O\,{\sc i}$)=13.82^{+0.09}_{-0.03}$ together with a
Doppler parameter of $11.9\pm3.5$. For gas with $0.1-1.0$
solar oxygen abundance the measured O\,{\sc i} column density corresponds
to an H\,{\sc i} column density range of log $N$(H\,{\sc i}$)=17.13-18.13$.
These numbers imply that this absorber is a LLS.
There is strong C\,{\sc iv} and Si\,{\sc iv} absorption in this 
system and FUSE data of PG\,1116+215 shows the presence of associated O\,{\sc vi}
absorption in the range $140-230$ km\,s$^{-1}$ with 
log $N$(O\,{\sc vi}$)=14.0$ (see Ganguly et al.\,2005 for details).
This high-velocity absorber therefore is also classified as
highly-ionized HVC with a dense cloud core that is responsible
for the O\,{\sc i} absorption.
Ganguly et al.\,(2005) speculate that this multi-phase absorber 
is related to the Magellanic Stream.

\section{Physical Conditions in the Absorbers}

\subsection{Setup of Ionization Models}

To constrain important physical parameters 
such as density, thickness, and ionization fraction of
the high-velocity metal-line absorbers we have used the 
photoionization code Cloudy (v96; Ferland et al.\,1998), which
allows us to model the observed column density ratios of low
and high ions. From ionization 
models for our halo absorbers it is immediately apparent
that the observed column densities 
of the weakly ionized species {\it together} with the column densities of 
the high ions cannot arise from a single gas phase.
Instead, the modelling implies that the low ions such as O\,{\sc i} and Si\,{\sc ii}
live in a region that is spatially separated from the gas phase in
which most of the high-ion absorption (C\,{\sc iv} and Si\,{\sc iv})
is arising (see also \S2). A plausible scenario is that 
the low ions populate a relative compact cloud core that is surrounded by a more 
extended, diffuse gas layer that hosts primarily intermediate
and high ions. Note that geometries such as sheet-like 
and/or filamentary structures are plausible for weak Mg\,{\sc ii} absorbers/sub-LLS (see, e.g., 
discussion in Milutinovi\a'c et al.\,2006). However, recent high-resolution 
21cm radio observations of Galactic halo LLS (Richter et al.\,2005;
Ben Bekhti et al.\,2009; in preparation) favour clump-like 
geometries with core-envelope structures for these systems.

{\it In further analysis we concentrate on the physical properties of
the cloud cores that are sampled by the low-ion absorption.}
The Si\,{\sc ii} and O\,{\sc i}
column densities listed in Table 2 have been derived from 
well-defined, unsaturated absorption lines, so that the Si\,{\sc ii}/O\,{\sc i}
ratios derived from these values are very reliable.
Both ions predominantly live in the same gas phase and in the
density range considered here, the Si\,{\sc ii}/O\,{\sc i}
ratio therefore can be used to constrain the physical
conditions in the gas.
We have constructed a grid of Cloudy ionization models 
to investigate for a given neutral hydrogen column density
and metallicity the dependence of the  Si\,{\sc ii}/O\,{\sc i}
column density ratio on the ionization parameter $U$, 
the ratio between ionizing photon density
and total particle density (i.e., $U=n_{\gamma}/n_{\rm H}$).
For an assumed ionizing radiation field one can calculate
$n_{\gamma}$ and thus estimate the gas density $n_{\rm H}$.
In our Cloudy model, absorbers are treated as plane-parallel
slabs with neutral hydrogen column densities varying from
log $N$(H\,{\sc i}$)=15.0$ to $19.5$ in steps of $0.5$ dex
(note that this version of Cloudy unfortunately cannot model other, 
more realistic absorber geometries).
In addition, we have used the Milky Way 
radiation field model, presented by Fox et al.\,(2005), to 
obtain the shape and the intensity of the ionizing flux as a 
function of the distance $d$ from the Galactic plane. The 
model is based on the theoretical work by 
Bland-Hawthorn \& Maloney (1999, 2002). It assumes that most
of the (hard) ionizing photons come from O-B stars in 
the spiral arms of the Milky Way and that the escape fraction
of hard photons into the halo is 6 percent (see Fox et al.\,2005
for a more detailed description).
The model predicts that the ionizing radiation field in the inner
$\sim 50$ kpc of the Milky Way is dominated by Galactic photons
from massive stars, while 
for larger distances the field is dominated by the extragalactic
UV background (Fox et al.\,2005, their Fig.\,6).
As an example of our Cloudy modeling, we show in Fig.\,6 
the predicted gas column densities for several atoms/ions 
assuming a neutral hydrogen column density of log $N$(H\,{\sc i}$)=17.0$,
a metallicity of $0.5$ solar, and a distance of the
absorber from the Galactic plane of $d=50$ kpc.

Our ionization models indicate that the Si\,{\sc ii}/O\,{\sc i}
column density ratio is a robust measure for the
ionization parameter for optically thin gas within the parameter range given
by the grid. Since we have $N$(O\,{\sc i}$)\gtrsim N$(Si\,{\sc ii})
in all our absorbers (see Table 3), the ionization
parameter $U$ can be estimated from $N$(O\,{\sc i}) 
and $N$(Si\,{\sc ii}) in a very simple manner (see also Fig.\,6):
  \begin{equation}
  {\rm log\,}U \approx 1.67 \left[ {\rm log\,}N({\rm Si\,II}) -
  {\rm log\,}N({\rm O\,I})+A_{\rm Si}\right] + B_U,
  \end{equation}
where $A_{\rm Si}$ denotes the depletion of Si into dust grains and 
$B_U$ is a scaling parameter that depends on the 
H\,{\sc i} column density. 
For low-column density halo clouds in the metallicity range $0.1-1.0$ solar 
one would expect values for $A_{\rm Si}$ in the
range $0.0-0.4$ (e.g., Richter et al.\,2001),
but not larger. For $B_U$, we find from the Cloudy calculations:
  \begin{equation}
  B_U \approx \left\{ \begin{array}{l@{\quad:\quad}l}
  -4.05 & 15.5 \leq {\rm log\,}N({\rm HI}) \leq 17.3 \\
  -4.05+0.71 \left[ {\rm log\,}N({\rm HI})-17.3 \right]
  & 17.3 < {\rm log\,}N({\rm HI}) \leq 18.5 \\
  \end{array} 
  \right\} .
  \end{equation}
Note that in our case $B_U$ depends neither critically 
on the metallicity of the gas
nor on the intensity or spectral shape of the assumed ionizing 
background (i.e., extragalactic versus Galactic ionizing radiation). 
From the neutral hydrogen column density together with 
the ionization parameter one finds for the total hydrogen
column density $N$(H)$=N$(H\,{\sc i}+H\,{\sc ii}):
  \begin{equation}
  {\rm log\,}N({\rm H}) \approx {\rm log\,}N({\rm HI})
  + 0.88\,{\rm log\,}U + 4.9.
  \end{equation}
Consequently, the hydrogen ionization fraction in 
optically thin gas, here defined as 
$f_{\rm H}=N$(H\,{\sc i}+H\,{\sc ii})/$N$(H\,{\sc i}), can
be expressed in terms of the ionization parameter in the form:
  \begin{equation}
  {\rm log\,}f_{\rm H} \approx 0.88\,{\rm log\,}U + 4.9.
  \end{equation}
Since the density of ionizing photons in the cloud, $n_{\gamma}$,
increases with decreasing distance $d$ to the Milky Way
disk, the relation between the ionization parameter $U$
and the hydrogen volume density $n_{\rm H}$ also is a 
function of $d$. Using the model of
Fox et al.\,(2005) together with our Cloudy grid we find that  
  \begin{equation}
  {\rm log\,}\,n_{\rm H} \approx
  {\rm log\,}X_{\gamma}(d) - {\rm log\,}U - 6.4,
  \end{equation}
where the parameter $X_{\gamma}(d)\approx 1,6$ and $35$ for cloud distances
of $d\geq100$ kpc, $d=50$ kpc, and $d=10$ kpc, respectively
(see Fox et al.\,2005, their Fig.\,6).
From this, the thickness of the absorbing gas layer $L$ 
(or cloud diameter in case of a spherical symmetry) can be obtained via:
  \begin{equation}
  L=\frac{N({\rm HI})+N({\rm HII})}{n_{\rm H}}.
  \end{equation}
  \noindent
Finally, under the assumptions that the absorber has 
spherical symmetry and that the sight line passes right
through the center of the spherical gas cloud with radius 
$r=L/2$, the mass of the cloud is given by
  \begin{eqnarray}
  M = \frac{4}{3} \, \pi \left( \frac{L}{2} \right) ^3 
  n_{\rm H}\, \mu \, m_{\rm H} \approx \\ \approx
  1.7\times 10^4 \, \left( \frac{n_{\rm H}}{{\rm cm}^{-3}}\right)
  \,\left( \frac{L}{\rm pc} \right)^3\,M_{\sun},
  \end{eqnarray}
where $\mu=1.4$ corrects for the presence of helium and heavy elements 
in the gas. However, even if our absorbers would have a roughly spherical
geometry, the chance that a random sightline passes through the 
geometrical center of the cloud is small (see, e.g., the case of Ton\,S210; 
Sembach et al.\,2002). Since the mass estimate depends on $L\leq2r$ to 
the third power we refrain from further using equation (8) 
to estimate cloud masses because of the resulting extremely large systematic 
uncertainties.

Equations (1)$-$(6) allow us to derive important physical quantities 
for the absorbing gas structures from the measured O\,{\sc i} and
Si\,{\sc ii} column densities assuming photoionization equilibrium,
a given spectral shape of the ionizing radiation, a given distance 
from the Milky Way, and a certain degree of dust depletion.  

\subsection{Physical Properties}

Using the formalism described above we have determined physical
conditions in the low-ion halo absorbers from the 
observed Si\,{\sc ii}/O\,{\sc i} column density ratios (Table 3).
Values for log [$N$(Si\,{\sc ii})/$N$(O\,{\sc i})] in our absorbers 
all lie in small range between $+0.03$ and $-0.65$, already suggesting that
the physical conditions are not vastly different. Assuming that 
the dust-depletion parameter for Si is $A_{\rm Si}=0.0-0.4$ (see 
previous section) one can obtain from equations (1) and (2) values for the
ionization parameter, $U$, and the ionization fraction, log $f_{\rm H}$, 
as a function of the neutral gas column density, $N$(H\,{\sc i}).
To further estimate volume densities and absorber thicknesses of
our clouds we have to assume a distance of the absorbing structures
to the disk of the Milky Way and its ionizing radiation (see equation 5).

We define two cloud models (model I and model II) with the same
metallicities (which scale with $N$(H\,{\sc i})) 
and silicon dust depletions, but with different assumptions for the distances.
Models I and II 
both assume a metallicity of $0.5$ solar and a zero depletion of Si 
into dust grains ($A_{\rm Si}=0$; see also \S4.4). The cloud distances to the Galaxy 
are set to $d=50$ kpc and $d=10$ kpc for models I and II, respectively.
As shown in Table 3, the distance-independent parameters $U$ and $f_{\rm H}$ 
for our seven absorbers range between log $U=-3.91$ to $-5.14$ and 
log $f_{\rm H}=0.38-1.46$. These numbers indicate that - regardless of 
the exact location of the absorbers in the halo - the cloud cores are 
substantially ionized with neutral hydrogen fractions ranging 
from $\sim 4$ to $\sim 40$ percent.
The hydrogen volume density range in our absorbers 
is $n_{\rm H}\approx 0.02-0.3$ cm$^{-3}$
in model I ($d=50$ kpc) and $0.1-2$ cm$^{-3}$ in model II ($d=10$ kpc). 
Absorber thicknesses are found to be $L\approx 0.2-130$ pc in model I and
$0.03-20$ pc in model II. Physical parameters for each absorber based 
on models I and II are listed in detail in Tables 4 and 5, respectively.

While it is not likely that the absorbers have a uniform metal abundance
or are located at the same distance to the disk, the results for 
models I and II provide useful information about the overall physical conditions
and characteristic dimensions of these structures.
The fact that we have detected seven weak high-velocity O\,{\sc i} absorbers
without 21cm counterparts in the LAB survey along a total number of 26 low-redshift
QSO sight lines with comparable S/N, suggests that the surface filling factor of these
absorbers is at least 25 percent, which is roughly twenty-five times the value found
for compact high-velocity clouds (Putman et al.\,2002). 

It is interesting to estimate the possible contribution of such a population of 
LLS and sub-LLS in the circumgalactic environment of the Milky Way to the 
gas-accretion rate of the Milky Way, assuming that the gas sooner or later
falls toward the disk. The gas accretion rate is given by 
$\dot{M}\approx M_{\rm LLS}\,\bar{v}/d$, where
$M_{\rm LLS}$ is the total mass of a population of LLS and sub-LLS gas clouds 
at a galactocentric radius $d$, and $\bar{v}$ is their mean infall velocity.
$M_{\rm LLS}$ can be estimated from the mean total gas column density 
$N$(H) and the surface filling factor $f$ via
 \begin{equation}
   M_{\rm LLS} \approx 10^7\,M_{\sun} \,f\,\mu \,
   \left(\frac{d}{10\,{\rm kpc}} \right)^2 \,
   \left(\frac{N({\rm H})}{10^{18}\,{\rm cm}^2} \right).
 \end{equation}
Assuming $N$(H$)=10^{18}$ cm$^{-2}$, $f=0.25$, and $\bar{v}=100$ km\,s$^{-1}$ 
we find for $d=10$ kpc values of $M_{\rm LLS} \sim 4\times 10^6 M_{\sun}$ 
and $\dot{M}\approx 0.04 M_{\sun}$ yr$^{-1}$. This is $\sim 10$ percent of the accretion
rate expected for the large HVC Complex C (Wakker et al.\,2008). 
However, if the LLS and sub-LLS were located at $d=100$ kpc the
total gas mass and the gas-accretion rate would be increased to
$M_{\rm LLS} \sim 4\times 10^8 M_{\sun}$ and 
$\dot{M}\approx 0.4 M_{\sun}$ yr$^{-1}$, thus of the same order
of magnitude as for the population of high-column density 21cm HVCs.

The typical angular diameter $\alpha$
of an individual spherical absorber with the diameter 
$L$ at a galactocentric distance $d$ is simply given by
$\alpha \,$[arcsec]$\approx 206\,L$\,[pc]$/d$\,[kpc].
Thus, for a characteristic cloud size of 1 pc (see
Tables 4 and 5) one finds $\alpha \approx 21$ arcsec for 
$d=10$ kpc and $\alpha \approx 2$ arcsec for $d=100$ kpc.
Such small angular sizes together with the surface filling
factor of $f=0.25$ would imply that there are more than
$10^8$ such "gas wisps" located in circumgalactic environment
of the Milky Way. Clearly, these numbers need to be re-evaluated
based on a larger sample of absorbers. However, also
recent hydrodynamical models of warm gas accretion onto the
Milky Way predict the presence of a huge number 
of small neutral and partly ionized gas fragments in
the Galactic halo (Bland-Hawthorn 2008).
Note that if the absorbers have a sheet-line geometry rather than
a spherical shape their number
could be substantially smaller while the filling 
factor would stay constant (see, e.g., Milutinovi\a'c et al.\,2006). Moreover, 
it seems possible that many of these small-scale structures group themselves
into larger gas complexes, similar as 
seen in the high-column density HVCs which are composed of
a large number of sub-structures (e.g., Sembach et al.\,2004). This would,
for instance, explain the presence of multiple velocity components 
in most of the absorbers.

Although the above considerations 
are based on various assumptions and low-number 
statistics and thus have to be taken with caution,
our study suggests that low-column density halo absorbers 
may be important for our understanding
of the complex gas-circulation processes in the circumgalactic
environment of the Milky Way. It is therefore desirable
to further investigate the frequency and physical properties of the halo LLS and sub-LLS
in a larger UV data set of extragalactic background sources. Such data 
hopefully will be provided by the upcoming {\it Cosmic Origins Spectrograph} (COS)
that will be installed on HST during Service Mission 4 in 2009.

\subsection{Metallicity}

The best indicator of the overall metallicity of neutral gas absorbers
is the O\,{\sc i}/H\,{\sc i} ratio because of the identical ionization 
potentials of neutral oxygen and neutral hydrogen and a strong 
charge-exchange reaction that couples these atoms in the neutral
interstellar medium.
As mentioned earlier, directly deriving H\,{\sc i} column densities
from the H\,{\sc i} Lyman series absorption is a difficult task because of 
the various blending problems in the lower and higher Lyman series lines 
and the often limited data quality. 
However, for the halo absorber (CHVC)
at $-167$ km\,s$^{-1}$ towards Ton\,S210 information about the H\,{\sc i} 
column density is available from both high-resolution 21cm observations with Parkes 
and FUSE absorption-line coverage of the H\,{\sc i} Lyman series 
(Sembach et al.\,2002). Sembach et al.\,derive a value of 
log $N$(H\,{\sc i}$)=18.34\pm0.04$; this, together with the neutral oxygen
column density of log $N$(O\,{\sc i}$)=14.02\pm0.04$ derived from
our STIS data, implies an oxygen abundance of 
[O/H]$=$log (O/H)$-$log (O/H)$_{\sun}=-1.01$ or $Z=0.098$ solar
((O/H)$_{\sun}=-3.31$; Asplund et al.\,2004). With an oxygen abundance of
$\sim 0.1$ solar this halo gas clump thus would have a metallicity 
similar to what has been found for the high-column density HVC
Complex C (e.g., Wakker et al.\,1999; Richter et al.\,2001; Sembach et al.\,2004).
Gas with such a low metallicity most likely does not originate
in the Milky Way but rather represents material that is being accreted by the
Milky Way from the intergalactic medium or from satellite galaxies. 
It has to be noted, however, that this result is affected by substantial
systematic uncertainties due to the H\,{\sc i} column density derived
from the beam-smeared 21cm data and the H\,{\sc i} Lyman series absorption
(see also discussion in Sembach et al.\,2002). The STIS data
clearly show the presence of at least two additional C\,{\sc ii} 
absorption sub-components blueward of the O\,{\sc i} absorption at $-167$ km\,s$^{-1}$ 
and one additional strong absorption component at $-242$ km\,s$^{-1}$ 
seen in intermediate and high ions (see $\S3.4$). This means that 
part of the H\,{\sc i} absorption seen in the unresolved lines of the 
Lyman series is not related to the O\,{\sc i} absorption, so that 
the true O\,{\sc i}/H\,{\sc i} ratio is underestimated.
Unfortunately, the FUSE data of Ton\,S210 are not accurate enough to reliably reconstruct
such a complex absorption pattern in the H\,{\sc i} Lyman series, so that
a quantitative estimate of the H\,{\sc i} column density 
in the satellite absorption components is not possible. We thus conclude
that the $0.1$ solar metallicity derived above can be regarded only as a 
{\it lower limit} on the actual metal abundance of the halo LLS in front of
Ton\,S210.

Another interesting aspect of the chemical composition of the LLS toward 
Ton\,S210 is the relatively large column density of Fe\,{\sc ii}. We measure
log $N$(Fe\,{\sc ii}$)=13.95\pm 0.06$, which $\sim 0.3$ dex above the 
column density of Si\,{\sc ii} (see Table 2). Our ionization modeling (Fig.\,6) shows
that this cannot be an ionization effect. On the contrary, the modeling
suggests that for a solar Si/Fe abundance ratio the Fe\,{\sc ii} column density
should be $\sim 0.1-0.4$ dex smaller than the Si\,{\sc ii}  column density 
for log $U<-3.9$ (see, e.g., the sub-LLS toward NGC\,4151; Table 2).
The large Fe\,{\sc ii}/Si\,{\sc ii} ratio in the absorber 
towards Ton\,S210 therefore suggests a substantial overabundance of
iron in this cloud and points toward a metal-enrichment of the 
gas from Type Ia supernovae (SNe). Interestingly, iron overabundances are
also observed in some of the intervening 
weak Mg\,{\sc ii} systems at $z\sim 1$ (Rigby et al.\,2002).

We have also searched for supplementary 
FUSE data in the MAST archive to explore the 
possibility of determining H\,{\sc i} column densities and metallicities 
for the other absorbers along the remaining five sightlines. 
From the available FUSE data only the spectra of PG\,1211+143 
and PG\,1116+215 have S/N sufficient to allow us to consider a
fit of the higher Lyman series lines at high radial velocities. 

We have re-reduced the FUSE spectrum of PG\,1211+143 
using the CALFUSE reduction pipeline version 3.2.0 (see Dixon et al.\,2007)
and have binned the data to $10$ km\,s$^{-1}$ pixel elements to improve the S/N.
Fig.\,7 shows the absorption profiles of six higher Lyman-series lines in the 
FUSE spectrum of PG\,1211+143 in the SiC\,1A channel in the LSR velocity
range between $-200$ to $+400$ km\,s$^{-1}$. Clearly,
high-velocity H\,{\sc i} absorption is seen in all these lines 
near $+174$ km\,s$^{-1}$ (in Fig.\,7 indicated with the dashed line), thus at
the same velocity at which high-velocity metal-absorption is observed 
in the STIS data (see $\S3.1$). However, since all these lines are fully
saturated and thus lie on the flat part of the curve-of-growth, the determination
of a reliable H\,{\sc i} column density from these data alone is very difficult.
To investigate qualitatively in what way the absorption {\it depths} 
depend on the H\,{\sc i} column density
we have created model H\,{\sc i} spectra for these lines using the FITLYMAN
routine implemented in the ESO MIDAS data analysis package (Fontana \& Ballester 1995).
For the model setup we have used a line FWHM that resembles the FUSE line-spread 
function in the SiC\,1A channel. 
As Doppler parameter we have chosen $b=12.0$ km\,s$^{-1}$,
which represents the best fit to the observed line {\it widths} in combination with the
assumed FWHM. The red lines in Fig.\,7 show the model spectra for H\,{\sc i} column
densities of log $N$(H\,{\sc i}$)=16.41,16.71$ and $17.41$. Together with 
the measured log $N$(O\,{\sc i}$)=13.10$, these H\,{\sc i} column densities 
translate to metallicities of $1.0,0.5$ and $0.1$ solar, respectively.
Note that the model with log $N$(H\,{\sc i}$)=17.41$ corresponds to the 
synthetic spectrum with the largest absorption depth. Although the 
FUSE data are relatively noisy, it is evident
that higher Lyman lines at $\lambda<930$ \AA\, never reach the zero-flux level in the 
high-velocity component but remain at a relative flux level of $\sim 0.2-0.3$. Only the model 
with log $N$(H\,{\sc i}$)=16.41$ reproduces this trend, while the two other models 
predict all displayed Lyman lines to reach the zero-flux level in the line cores. Note 
that the observed residual flux in the line center cannot be caused by any kind of background in
the FUSE data since the nearby strong local H\,{\sc i} absorption at zero velocities
does indeed go down to zero intensities. 
We conclude that log $N$(H\,{\sc i}$)\approx16.41$ most likely 
represents a realistic estimate for the total neutral hydrogen column density
in the high-velocity halo gas near $+191$ km\,s$^{-1}$, 
while larger H\,{\sc i} column densities are not supported by the
FUSE data. 

As with Ton\,S210 there is a systematic problem in reliably 
estimating the metallicity of the sub-LLS toward PG\,1211+143 because of the presence
of sub-component structure, as
seen in the lines of C\,{\sc ii}, Si\,{\sc iii}, and C\,{\sc iv} 
near $+193$ km\,s$^{-1}$ towards PG\,1211+143 (\S3.1). 
Although this satellite 
component apparently is very weak, it does 
contribute to the high-velocity H\,{\sc i} Lyman series absorption, 
but cannot be resolved in the noisy FUSE data. Taking this into account, 
the O\,{\sc i}/H\,{\sc i} column-density 
ratio is $\geq 4.9 \times 10^{-4}$, corresponding to [O/H$]\geq 0.00$. 
Therefore, this absorber apparently 
has a metallicity of solar or slightly above solar,
suggesting that the gas is not extragalactic but 
originates in the Galactic disk, possibly as part
of a Galactic fountain flow or a similar process that
circulates metal-enriched 
material from the Galactic plane into the halo and back.

In a very similar way, Ganguly et al.\,(2005) previously have 
determined the metallicity of the LLS at $+193$ km\,s$^{-1}$ 
towards PG\,1116+215 using FUSE data. 
They derive an H\,{\sc i} column density 
of log $N$(H\,{\sc i}$)=17.82^{+0.12}_{-0.14}$
in this cloud. Together with the measured O\,{\sc i} column
density (log $N$(O\,{\sc i}$)=13.82$; see \S3.6) the metallicity
is $\sim 0.2$ solar, assuming that O\,{\sc i} and all of the H\,{\sc i}
reside in the same gas phase.
However, because of the complex multi-phase absorption of this system,
part of the H\,{\sc i} Lyman series absorption
most likely is related to the highly-ionized gas phase traced by the
intermediate and high ions. We therefore 
regard the oxygen abundance
of $\sim 0.2$ solar for the high-velocity LLS towards PG\,1116+215 
as a lower limit for its metallicity.

Metal-abundance analyses of three of the low-column density absorbers 
show that their gas is metal-enriched at levels between ten percent solar and
roughly solar. This is an abundance range similar to that of the 
high-column density HVCs and IVCs (e.g., Richter et al.\,2001).
Such metallicities suggest that both 
Galactic and extragalactic material circulates through the halo
of the Milky Way and produces numerous metal absorbers that 
exhibit a large range in column density and linear size. The 
overabundance of iron in one of the clouds further implies that
both stellar explosion mechanisms (Type Ia and Type II SNe) 
must have contributed to the metal budget in the gas.

\subsection{Dust content}

In the neutral interstellar medium (ISM) 
calcium is nearly always found to be depleted 
into dust grains (e.g., Crinklaw, Federman \& Joseph 1994).
As a result, measured Ca\,{\sc ii}/H\,{\sc i} 
and Ca\,{\sc ii}/O\,{\sc i} column density ratios are
much lower (more than 1.5 dex, typically) 
than what would be expected from solar 
calcium-to-hydrogen and calcium-to-oxygen
abundance ratios. Moreover,
the ionization potentials of Ca\,{\sc ii} is $11.87$ eV (Morton 2003),
thus less than the ionization potential of neutral
hydrogen and neutral oxygen ($13.6$ eV). 
In low-density regions in the neutral ISM
the dominant ionization state of calcium thus is 
Ca\,{\sc iii} rather than Ca\,{\sc ii}. 
Ca\,{\sc ii}/H\,{\sc i} and Ca\,{\sc ii}/O\,{\sc i}
column density ratios consequently are further reduced
in diffuse, neutral gas clouds by ionization effects.

Ca\,{\sc ii} and O\,{\sc i} absorption
has been detected in the two high-velocity LLS at
$-207$ and $-163$ km\,s$^{-1}$ towards PHL\,1811 (see \S3.5). This finding 
allows us to evaluate possible dust depletion
of calcium in these absorbers.
Assuming a solar calcium abundance of (Ca/H$)_{\sun}=-5.65$ (Morton 2003)
and ignoring any ionization corrections, the measured Ca\,{\sc ii}/O\,{\sc i}
column-density ratios imply [Ca/O$]=-0.02$ for the absorber
at $-207$ km\,s$^{-1}$ and [Ca/O$]=+0.24$ for the system at $-163$
km\,s$^{-1}$. Within the $2\sigma$ error range both ratios therefore
are consistent with a solar calcium-to-oxygen abundance ratio.
This is a remarkable result. First, it indicates that Ca\,{\sc ii}
- despite the relatively low ionization potential - is
the dominant ionization state of calcium in these high-velocity
LLS and sub-LLS. If significant amounts of calcium in the 
O\,{\sc i}/H\,{\sc i} phase was in the form of
Ca\,{\sc iii}, then the Ca/O abundance ratio would be
considerably above solar, which seems unlikely.
Second, the derived solar [Ca/O] ratios imply that there obviously is
{\it no depletion of calcium into dust grains} in these
absorbers. For 
comparison, Savage \& Sembach (1996) investigated in detail the
dust depletion pattern of various heavy elements in Milky Way disk and halo 
gas towards the halo star HD\,116852. For their "warm halo clouds" they
find Ca\,{\sc ii}/Zn\,{\sc ii} ratios that are more than 1.6 dex below
the solar Ca/Zn ratio, showing that that 
a substantial amount of calcium is locked into dust grains. 
In striking contrast, the solar
Ca/O abundance measured for the two high-velocity absorbers towards
PHL\,1811 strongly suggests that these gas clouds are basically 
dust-free.

A plausible explanation for the lack of dust in these clouds is that
the dust grains have been destroyed by the same energetic event that
has driven out the gas from the galaxy of origin 
into the Milky Way's circumgalactic environment.
Additional observations of Ca\,{\sc ii} in the other high-velocity LLS and
sub-LLS are required to determine whether the lack of dust depletion
is typical for these kind of absorbers or whether this is an
unusual characteristic of the two absorbers towards PHL\,1811.

\section{Discussion}

Our study has unveiled a 
population of weak low-ion absorbers at high radial velocities that presumably 
are located in the halo of the Milky Way and/or its immediate intergalactic 
environment. 
These circumgalactic gas wisps may simply represent the low-column density tail
of common HVCs, but they are obviously also somehow connected to the 
population of compact HVCs (CHVCs) and highly-ionized HVCs. While it is unclear
whether these absorption systems contribute significantly to the total gas mass 
in the halo of the Milky Way, they obviously do have a substantial absorption
cross section in low and intermediate ions. It therefore is evident that these absorbers 
could represent a very important link between weakly and highly-ionized and
low- and high-mass gaseous structures in the inner and outer halo of the Milky Way.
For these reasons it is of interest to compare
the physical and statistical properties of these systems 
with the characteristics of intervening metal absorbers 
at low and high-redshifts.

Absorption by neutral and weakly ionized species is commonly found in the extended
($< 100\,h^{-1}$ kpc) halos of galaxies, suggesting the presence of cool gas in the form
of clouds or filamentary structures that are embedded in the more tenuous
hot coronal gas. Important information about the column-density 
distribution and ionization conditions of the intergalactic
gaseous environment of galaxies comes from QSO absorption-line 
studies of intervening Mg\,{\sc ii} absorbers (e.g., Bergeron \&
Boiss\a'e 1991; Charlton \& Churchill 1998; Rigby et al.\,2002).
Strong Mg\,{\sc ii} absorbers ($W_{2796} > 0.3$ \AA) 
preferentially are associated with luminous galaxies 
($L>0.05\,L^{\star}$) at impact parameters $< 35\,h^{-1}$ kpc 
(e.g., Churchill et al.\,1999; Ding et al.\,2005; Kacprzak et al.\,2008), 
suggesting that high-column
density absorption (log $N$(H\,{\sc i})$>18$, typically)
takes place in the gaseous disks of these galaxies and/or
in the immediate circumgalactic environment (e.g., from intergalactic 
gas that is being accreted by these galaxies).
The weak Mg\,{\sc ii} systems ($W_{2796} \leq 0.3$ \AA), in contrast,
appear to have no such tight association with galaxies, but
are typically found at larger distances from luminous
galaxies, in the range $35-100\,h^{-1}$ kpc (Milutinovi\a'c et al.\,2006).
It is rather surprising that many weak Mg\,{\sc ii} systems 
appear to have high metallicities ($>0.1$ solar) and high
iron abundances (Rigby et al.\,2002). 
This possibly indicates that either "in situ" star formation is
responsible for their enrichment (Rigby et al.\,2002)
or that there exists a mechanism that efficiently
circulates metal-enriched material
from star-formation sites within galaxies to the outer
circumgalactic environment at large distances (e.g.,
"superwinds", Rosenberg et al.\,2003, or "intergalactic fountains").
Many of the metals that are found in absorption-line
systems around massive galaxies could
be produced by the ejecta of adjacent faint, dwarf galaxies (i.e.,
satellite galaxies). As an example, Stocke et al.\,(2004) 
have discovered a faint ($M_B=-13.9$), dwarf poststarburst galaxy that
is likely to be responsible for a sub-LLS with similar radial
velocity towards the nearby 
3C\,273 sightline. The angular separation between the dwarf galaxy
and the absorber implies a projected distance of 
$\sim 70$ $h_{70}^{-1}$ kpc at the Hubble-flow distance
of the galaxy ($\sim 23$ Mpc). Stocke et al.\,
argue that starbursting dwarf galaxies may be primaribly responsible
for intervening, weak metal-absorbers at low and high redshift. 
Such galaxies would be difficult to detect at larger distances because
they are so faint.

As weak metal-line absorption of low-ionization species such as Mg\,{\sc ii}
represents a common phenomenon along QSO sight lines that pass through
the outer halos of galaxies, one should expect that there exists
a corresponding population of weak absorbers nearby, i.e., within 
the gaseous environment of the Local Group.
The high-velocity LLS and sub-LLS discussed in this paper currently are
the most promising candidates for being the local counterparts of the weak
Mg\,{\sc ii} systems. If they are, one would expect that these absorbers
are located at rather large distances from the Galactic plane ($30-100$ kpc).
The relatively large radial velocities found for our seven LLS and sub-LLS
generally are consistent with such large distances.
However, since we do not have further information on the distances of the 
absorbers, a more nearby location cannot be excluded (see discussion below).
Note that in our photoionization modeling we have taken into account
this uncertainty, and have
derived distance-dependent physical parameters (see Tables 4 \& 5).

Our study indicates that high-velocity LLS and sub-LLS fill 
about 25 percent of the sky, implying that they are numerous.
However, to more reliably
evaluate statistical and physical similarities of these objects with
intervening weak Mg\,{\sc ii} systems at low redshift, 
a much larger UV data set of weak high-velocity O\,{\sc i}/Si\,{\sc ii}/Mg\,{\sc ii}
absorbers is required. Such data will be particularly important to 
study in detail the area filling
factors of the LLS/sub-LLS, the connection between low-ion absorbers
and high-ion systems, as seen in O\,{\sc vi} and C\,{\sc iv}
(see, e.g., Sembach et al.\,1999; 
Sembach et al.\,2003; Milutinovi\a'c et al.\,2006),
and their kinematics (e.g., Mshar et al.\,2007).
Fortunately, the COS instrument - after it has
been successfully installed on HST - is projected
to obtain new UV data of several dozen QSOs that can be used
for this purpose.

An important result of our study is the simultaneous detection 
of O\,{\sc i} and Ca\,{\sc ii} absorption in the two weak halo absorbers 
towards PHL\,1811. This finding demonstrates that
high-velocity Ca\,{\sc ii} absorbers trace neutral gas structures in the
halo with H\,{\sc i} column densities as low as log $N$(H\,{\sc i}$)=17.5$
(or even lower, if the metallicity is higher than $0.1$ solar; see
Table 3). This important finding links our previous studies on high-velocity
Ca\,{\sc ii} absorption in the Milky Way halo (Richter et al.\,2005;
Ben\,Bekhti et al.\,2008) with the UV-selected LLS and sub-LLS
presented in this study. The presence of Ca\,{\sc ii} absorption
in the high-velocity LLS and sub-LLS
suggests that the two Ca\,{\sc ii} H\&K lines near
$4000$ \AA\, may provide an excellent tool
to study the distribution of optically thick neutral gas
in the halos of the Milky Way and other galaxies at low redshift.
We therefore are planning to investigate the presence or
absence of optical Ca\,{\sc ii} absorption in LLS and sub-LLS
in the Milky Way halo along the other sightlines in our STIS sample,
using ground-based telescopes and high-resolution spectrographs.

As for the 21cm HVCs, one important question about the nature of the
high-velocity LLS and sub-LLS concerns the origin of the gas.
If the distance to these objects is $10$ kpc rather than $30-100$ kpc,
a Galactic origin as part of a galactic fountain (Shapiro \& Field 1976) 
or other outflow mechanisms seems possible, although the bulk of
galactic fountain material is expected to take form of 
solar-metallicity intermediate-velocity clouds located typically 
within $\sim 1$ kpc of the disk (e.g., Wakker et al.\,2008).
Alternatively, the gas may originate in the intragroup medium
of the Local Group or may represent material torn out of 
smaller satellite galaxies, as they are being accreted by the
Milky Way and M31. It is clear from previous HVC studies 
that all of the above mentioned processes contribute to 
the circulation of neutral and ionized gas in the circumgalactic 
environment of the Milky Way (see Richter 2006 for a recent review).
Thus, the origin of the high-velocity LLS and sub-LLS may be
manifold as well. 

Four of the halo LLS and sub-LLS are located near the northern 
Galactic pole in the range $l=155-291$ and $b=+64$ to $+75$. They
all have large positive radial velocities between $+145$ and $+293$ 
km\,s$^{-1}$. Given these parameters, a plausible origin for these
structures is the Magellanic Stream and its Leading Arm (LA; see Fig.\,1).
The LA, as seen in H\,{\sc i} 21cm emission, has high positive radial
velocities at $l>240$ and $b>0$, 
as indicated with the red colour in Fig.\,1. Low-column density
absorbers with similar radial velocities near the northern Galactic pole 
thus could represent the extension of the LA in this direction. Such
an extension may have the form of a coherent stream of scattered, 
partly ionized cloudlets whose individual H\,{\sc i} column densities
are below the detection limit of 21cm radio observations.
While the lower metallicity limit of $\sim 0.2$ solar 
for the high-velocity LLS towards PG\,1116+215 is fully consistent
with a LA origin (see also Ganguly et al.\,2005 for a detailed 
discussion on this issue), the nearly solar metallicity of the 
sub-LLS towards PG\,1211+143 speaks against a LA origin for
this absorber. Alternatively, these absorbers may represent
stripped or ejected gas from one or several dwarf galaxies in the Local
Group. Possible candidates are the two dwarf spheroidals Leo\,I and Leo\,II,
which are located at $l=226$, $b=+49$ and $l=220$, $b=+67$, respectively.
Both galaxies have high positive radial velocities in a range similar 
to that of the high-velocity absorbers.
Finally, these four positive-velocity absorbers could also represent
gas {\it outflowing} from the Milky Way disk. In directions near the
northern Galactic pole the contribution of the disk rotation to
the observed radial velocities should be very small. Consequently,
the high positive radial velocities of the absorbers suggest that the gas
is moving away from us rather than falling toward the disk.
Evidence for a Galactic outflow also comes from the 
observed high-velocity wings seen in O\,{\sc vi} absorption
profiles towards extragalactic background sources (e.g., 
Fox, Savage \& Wakker 2006).
Some of the high-velocity LLS and sub-LLS in the halo 
thus might represent high-density peaks in an outflowing 
stream of mostly hot, ionized gas. However, for the 
three systems located in the 
southern Galactic hemisphere,
the lack of dust depletion in the two systems
towards PHL\,1811 and the super-solar Fe/$\alpha$ ratio in the
cloud towards Ton\,S210 clearly favor an extragalactic 
origin.

In addition to the origin and distribution of the 
high-velocity LLS and sub-LLS there remain many
open questions about their physical nature.
Our Cloudy models imply characteristic ionization parameters of 
log $U<-3.9$ and typical hydrogen volume densities of 
$n_{\rm H}>0.02$ cm$^{-3}$ in the O\,{\sc i}/Si\,{\sc ii} 
absorbing regions. As these relatively dense (and  
for the expected low thermal gas pressure, presumably relatively cold) gas 
clumps move with considerable velocities through the ambient hot 
coronal gas, one would expect from hydrodynamical considerations
that they should be destroyed on relatively
small time scales by Kelvin-Helmholtz and Rayleigh-Taylor 
instabilities (e.g., Jones, Kang \& Tregillis 1994). 
Therefore, high-velocity LLS and sub-LLS in the Milky Way halo 
possibly represent {\it transient phenomena}, 
e.g., short-lived, local density enhancements 
within an infalling or outflowing stream of hot gas.
Such short-lived density enhancements are indeed observed in 
hydrodynamical simulations of Galactic winds (see review by 
Veilleux, Cecil \& Bland-Hawthorn 2005 and references therein).
Interestingly, Schaye et al.\,(2007) have found a population of possibly short-lived,
metal-rich C\,{\sc iv} absorbers at $z\approx 2.3$ using optical spectra from
VLT/UVES. These authors argue that the C\,{\sc iv} systems represent transient
structures that govern the metal-transport from star-forming galaxies into
the IGM in the form of small, metal-rich patches of gas. 
Moreover, Schaye et al.\,discuss the possibility
that high-metallcity C\,{\sc iv} absorbers, weak Mg\,{\sc ii} systems and
high-velocity Ca\,{\sc ii} absorbers in the Milky Way halo
represent the same circumgalactic absorber 
population at different evolutionary stages (see Schaye et al.\,2007, their \S6.5).

To further investigate these interesting aspects it would be of great importance
to exactly localize the LLS and sub-LLS in the circumgalactic environment
of the Milky Way, i.e., to determine their radial distribution and
thus their distances. However, direct distance measurements of such small
gaseous features in the halo will be almost impossible, because
it is extremely unlikely to find a set of 
appropriately aligned halo stars with known distances
over such small areas on the sky (see also Wakker et al.\,2007; 
Wakker et al.\,2008).
It is therefore important to also investigate the gas distribution
and dynamics
around galaxies and its absorption characteristics using
numerical simulations (e.g., Fangano et al.\,2007). This
is very challenging, too, given the extremely large
dynamic range that such simulations need to span 
(sub-pc scale clouds in a Mpc-scale galaxy halo) and 
the complex physics that needs to be included 
to derive realistic results 
(see also Fujita et al.\,2008).

The advent of fresh UV absorption-line data from COS
and the rapid improvement of numerical simulations of galaxies
and their environment raises hope, however, that we will soon
have a better understanding the complex distribution of neutral and 
ionized gas in the outskirts of the Milky Way and other 
galaxies. 

\section{Summary}

In this paper we have studied a 
population of weak, high-velocity metal-line absorbers, 
located in the halo and/or circumgalactic environment of 
the Milky Way, using archival high-resolution UV quasar spectra from
HST/STIS. \\
\\
1. Along six QSO sightlines towards PG\,1211+143, RJ\,J1230.8+0115,
   NGC\,4151, Ton\,S210, PHL\,1811, and PG\,1116+215  we identify seven weak 
   O\,{\sc i}/Si\,{\sc ii} absorption systems at radial velocities 
   $|v_{\rm LSR}|=100-320$ km\,s$^{-1}$. We measure O\,{\sc i} column 
   densities in these absorbers of $N$(O\,{\sc i}$)<2 \times 10^{14}$ cm$^{-2}$,
   corresponding to H\,{\sc i} column densities of $\sim 10^{16.2}-10^{18.4}$ cm$^{-2}$,
   assuming an oxygen abundance of $0.1-1.0$ solar.  These absorbers thus 
   represent Lyman-Limit systems and sub-Lyman-Limit systems (LLS and sub-LLS)
   with H\,{\sc i} column densities below the typical detection
   limits of all-sky 21cm HVC surveys.\\
\\
2. The detected high-velocity LLS and sub-LLS appear not be directly associated with 
   known large 21cm HVC complexes. However, in at least one case (Ton\,S210) 
   there is a physical connection between the high-velocity absorption and 21cm 
   emission from a compact high-velocity cloud (CHVC). In three cases 
   (Ton\,S210, PHL\,1811 and PG\,1116+215) the
   neutral gas absorption is associated with highly-ionized gas 
   seen in O\,{\sc vi} absorption at similar radial velocities. The high-velocity
   LLS and sub-LLS thus appear to be partly embedded in more tenuous, low-density
   gas, being part of the complex multi-phase circumgalactic gaseous envelope
   of the Milky Way.\\
\\
3. We have modeled the ionization conditions in the high-velocity LLS and sub-LLS
   and have obtained characteristic ionization parameters of log $U<-3.9$ and typical
   hydrogen densities of $n_{\rm H}>0.02$ cm$^{-3}$, based on the observed
   O\,{\sc i}/Si\,{\sc ii} column-density ratios. These numbers suggest that
   the absorbers have linear sizes on pc and sub-pc scales. With a filling 
   factor of $\gtrsim 25$ percent these structures appear to be very numerous
   ($N>10^8$) and their total mass (depending on the distance) may be
   comparable to the total mass of the 21cm HVCs.\\
\\
4. Using supplementary UV absorption data from FUSE and considering 
   previous results by Ganguly et al.\,(2005) we constrain the metallicity of three
   of the absorbers towards PG\,1211+143, Ton\,S210 and PG\,1116+215 to be 
   in the range $\sim 0.1-1.0$ solar. This is similar to what is found for 
   the large 21cm IVC and HVC complexes. Using supplementary optical absorption
   spectra from VLT/UVES we derive, for the two halo absorbers towards PHL\,1811,
   Ca\,{\sc ii}/O\,{\sc i} column-density ratios that correspond to
   solar Ca/O abundance ratios. The lack of dust depletion of calcium
   in these absorbers thus indicates that they do not contain significant amounts 
   of dust.\\
\\
5. Four of the high-velocity LLS and sub-LLS have high positive radial velocities
   and are located near the northern Galactic pole. These systems 
   may indicate scattered gas clouds belonging to a low-column density tail
   of the Leading Arm of the Magellanic Stream.
   Alternatively, these four absorbers may represent gas ejected from
   dwarf galaxies in the Local Group (e.g., Leo\,I and Leo\,II) or 
   gaseous structures that 
   participate in an outflow from the Milky Way disk as part of a Galactic wind.
   The high-velocity LLS and sub-LLS in the halo possibly 
   are transient phenomena, e.g., short-lived density enhancements 
   produced by instabilities in an outflowing and/or infalling stream of (hot) gas.\\
\\
6. We speculate that the high-velocity LLS and sub-LLS represent the local counterparts
   of intervening weak Mg\,{\sc ii} absorbers that are commonly found in quasar
   lines of sight that pass the circumgalactic gaseous environment of other galaxies.
   These absorption systems therefore may represent an important link between the 
   outer gaseous halo of the Milky Way and 
   metal-line absorbers in quasar spectra at low and high redshift. Future studies
   of high-velocity LLS and sub-LLS are of great importance to better
   understand the complex distribution of neutral and ionized gas in the outer
   regions of galaxies in the context of a hierarchical galaxy-evolution scenario.

\acknowledgments

P.R. and N.B.B. acknowledge financial support by the German
\emph{Deut\-sche For\-schungs\-ge\-mein\-schaft}, DFG,
through Emmy-Noether grant Ri 1124/3-1. J.C. was funded
by NASA under grant NAGS-6399 and by NSF under grant
AST04-07138.



\clearpage


\clearpage

\begin{figure}
\epsscale{1.0}
\plotone{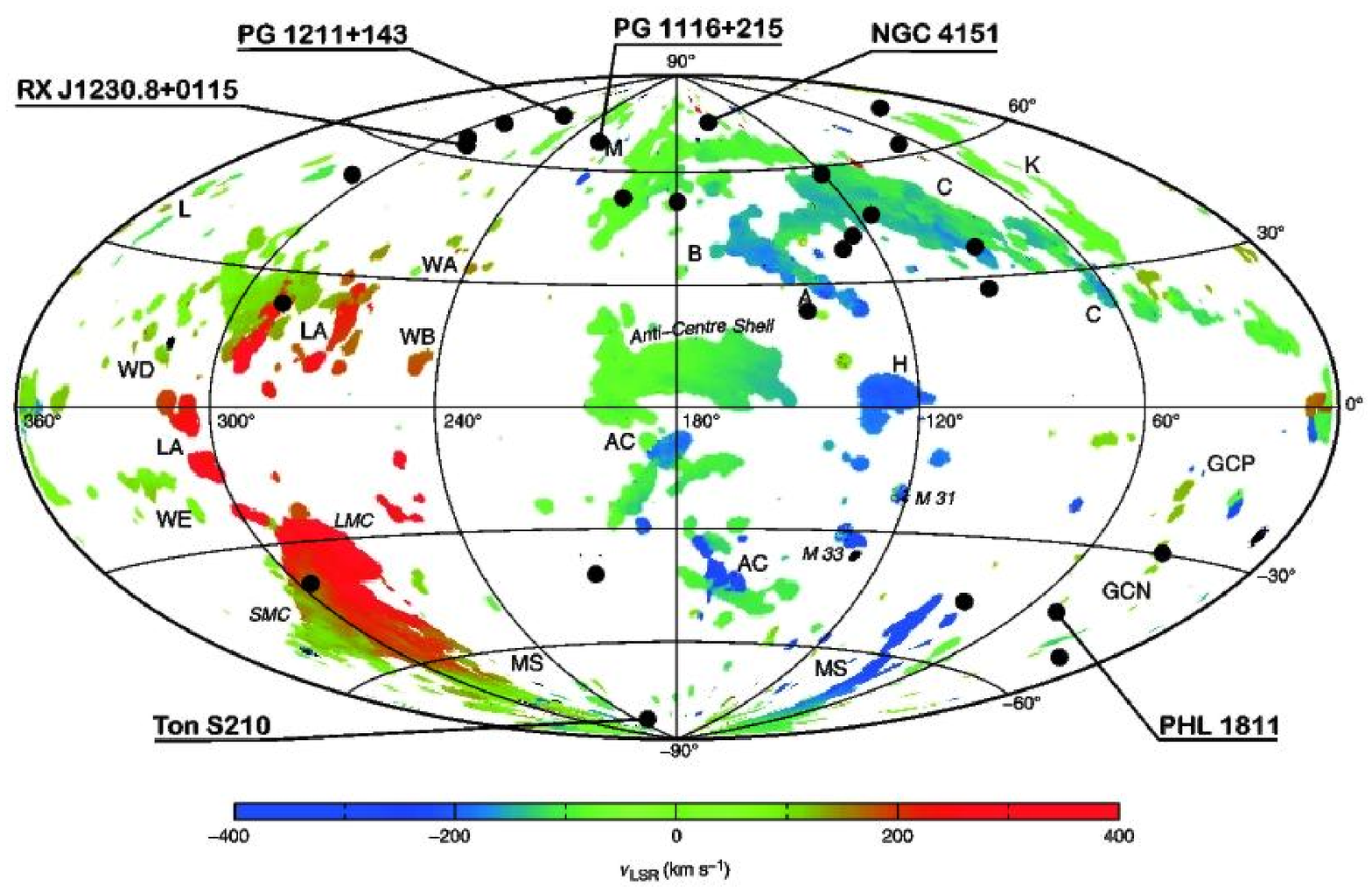}
\caption{
An H\,{\sc i} 21cm all-sky map of Galactic high-velocity clouds is shown, 
based on the data from the LAB survey (Kalberla et al.\,2005) and a model
of the Milky Way gas distribution (Kalberla et al.\,2007). The LSR radial 
velocities are color-coded according to 
the color scheme displayed as a bar in the lower
part of the plot. The position of the 26 QSO sightlines discussed in this
paper are indicated with filled dots. 
The LAB all-sky map was kindly provided by T. Westmeier. A HIGH-RESOLUTION
VERSION OF THIS FIGURE IS AVAILABLE ON REQUEST.
}
\end{figure}

\begin{figure}
\epsscale{1.0}
\plotone{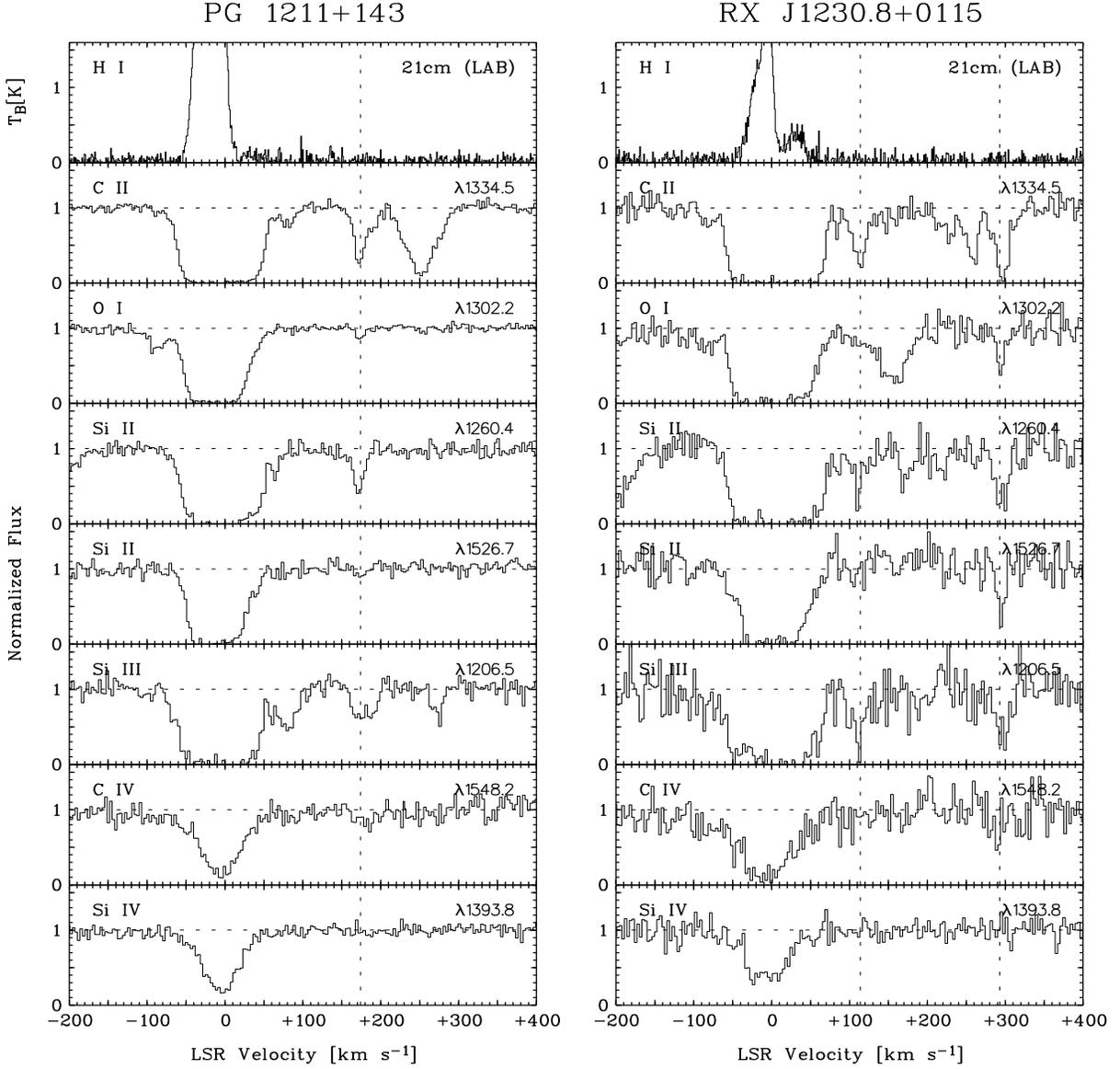}
\caption{
Continuum-normalized absorption profiles of low and high ions 
(STIS data) are plotted together with the 
H\,{\sc i} 21cm emission profiles (LAB data) for the lines of 
sight towards PG\,1211+143 and RX\,J1230.8+0115 
against the LSR radial velocity.
High-velocity absorption from gas in the Milky Way halo is
indicated with dashed lines. Note that the absorption
near $+250$ km\,s$^{-1}$ in the C\,{\sc ii} velocity restframe
belongs to Galactic absorption of C\,{\sc ii}$^{\star}$.
}
\end{figure}

\clearpage

\begin{figure}
\epsscale{1.0}
\plotone{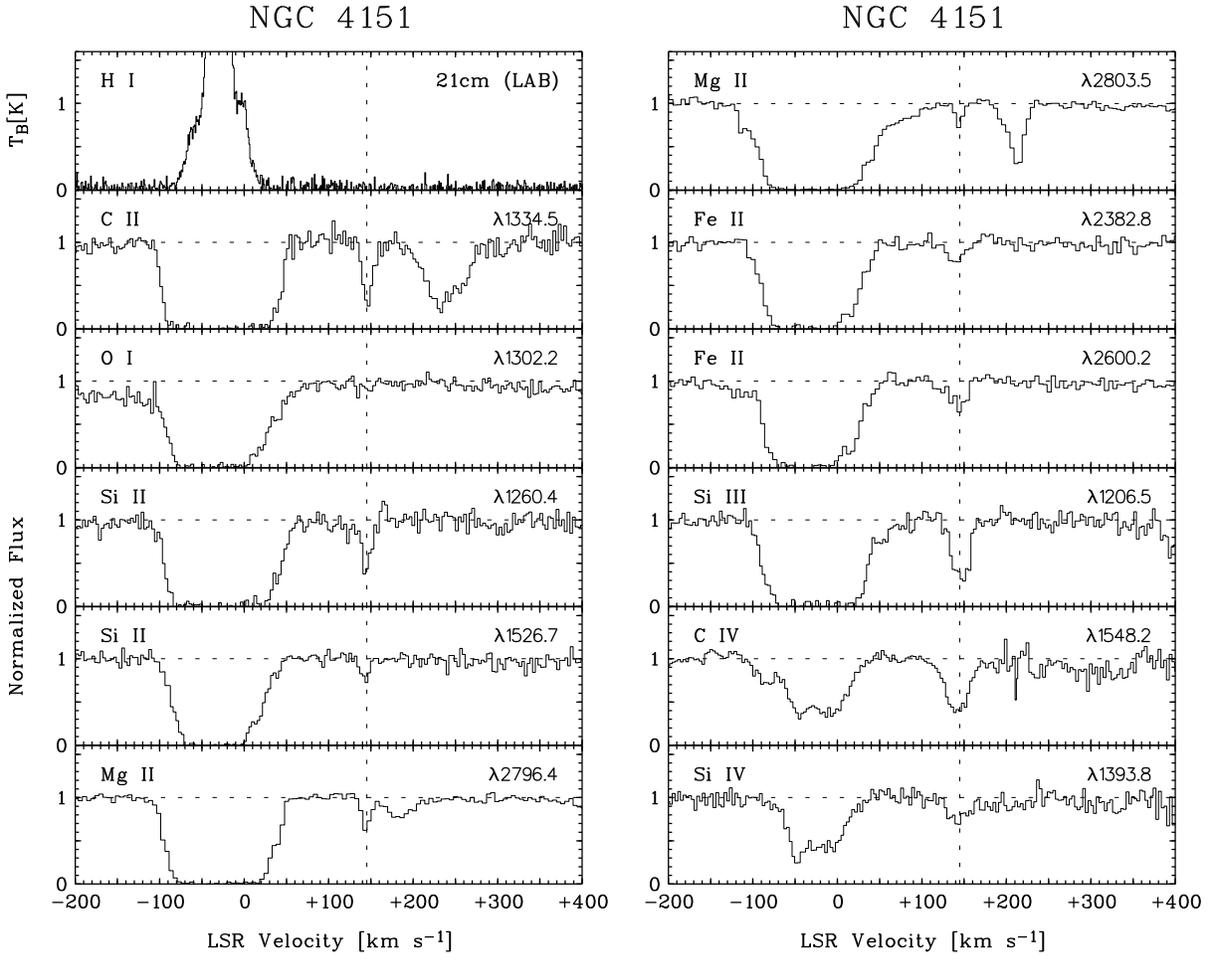}
\caption{
Same as Fig.\,2., but for the line of sight towards
NGC\,4151.
}
\end{figure}

\clearpage

\begin{figure}
\epsscale{1.0}
\plotone{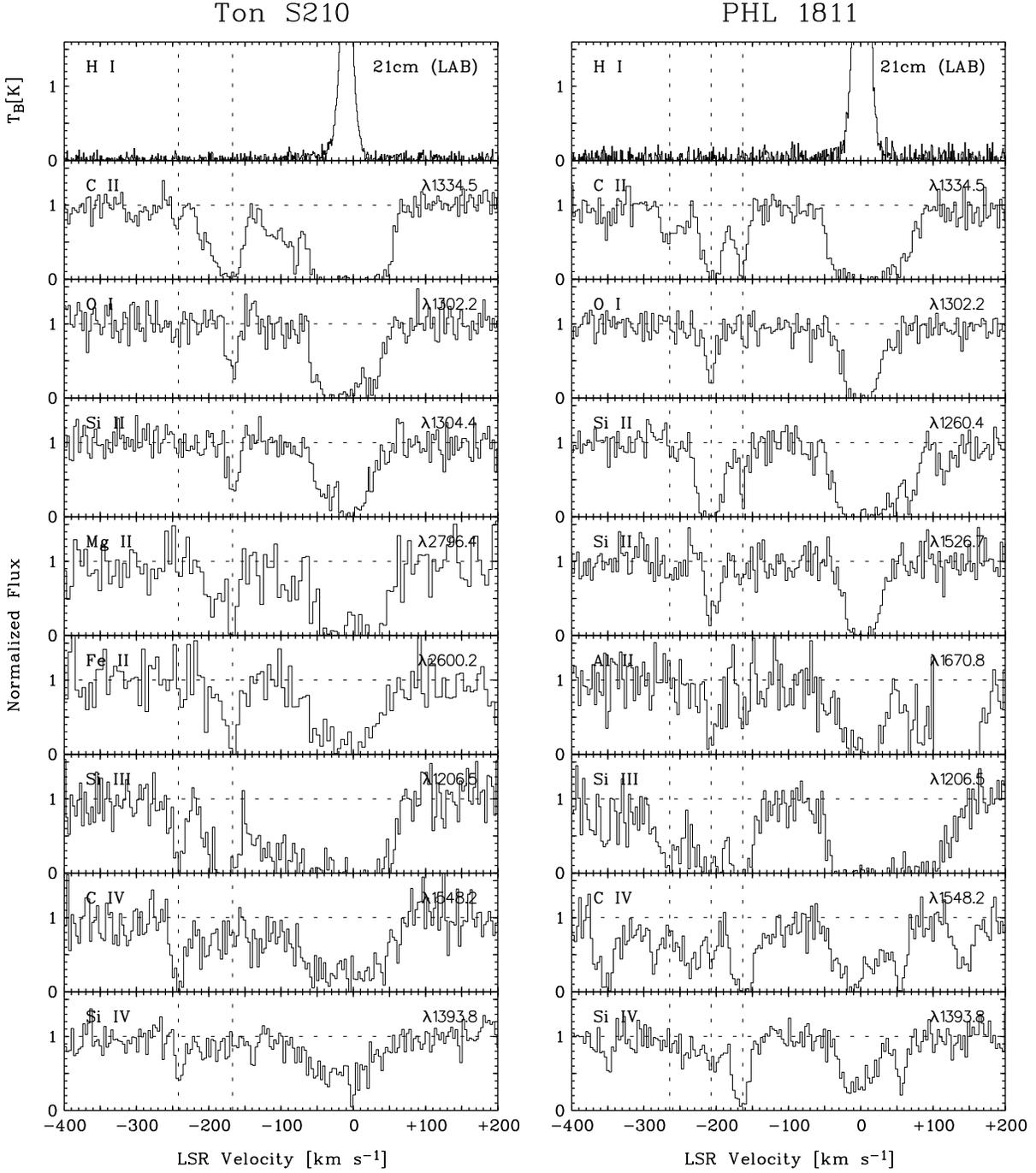}
\caption{
Same as Fig.\,2., but for the lines of sight towards
Ton\,S210 and PHL\,1811.
}
\end{figure}

\begin{figure}
\epsscale{0.6}
\plotone{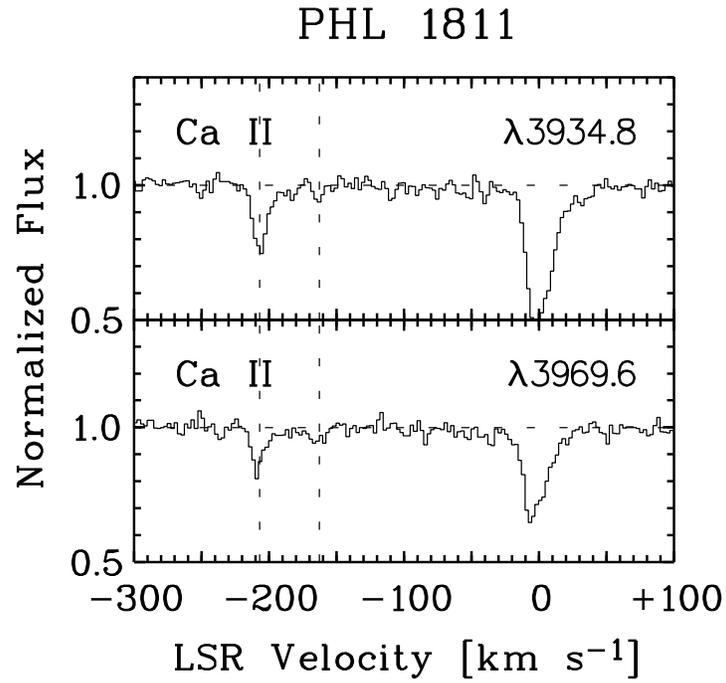}
\caption{
High-velocity Ca\,{\sc ii} absorption towards PHL\,1811, as observed with 
UVES/VLT (data adopted from Ben\,Bekhti et al.\,2008). The two high-velocity
Ca\,{\sc ii} absorption components at $-207$ and $-163$ km\,s$^{-1}$ (indicated
with dashed lines) coincide with
the two main high-velocity absorption components 
seen in the STIS data (see Fig.\,4).
}
\end{figure}

\clearpage

\begin{figure}
\epsscale{1.0}
\plotone{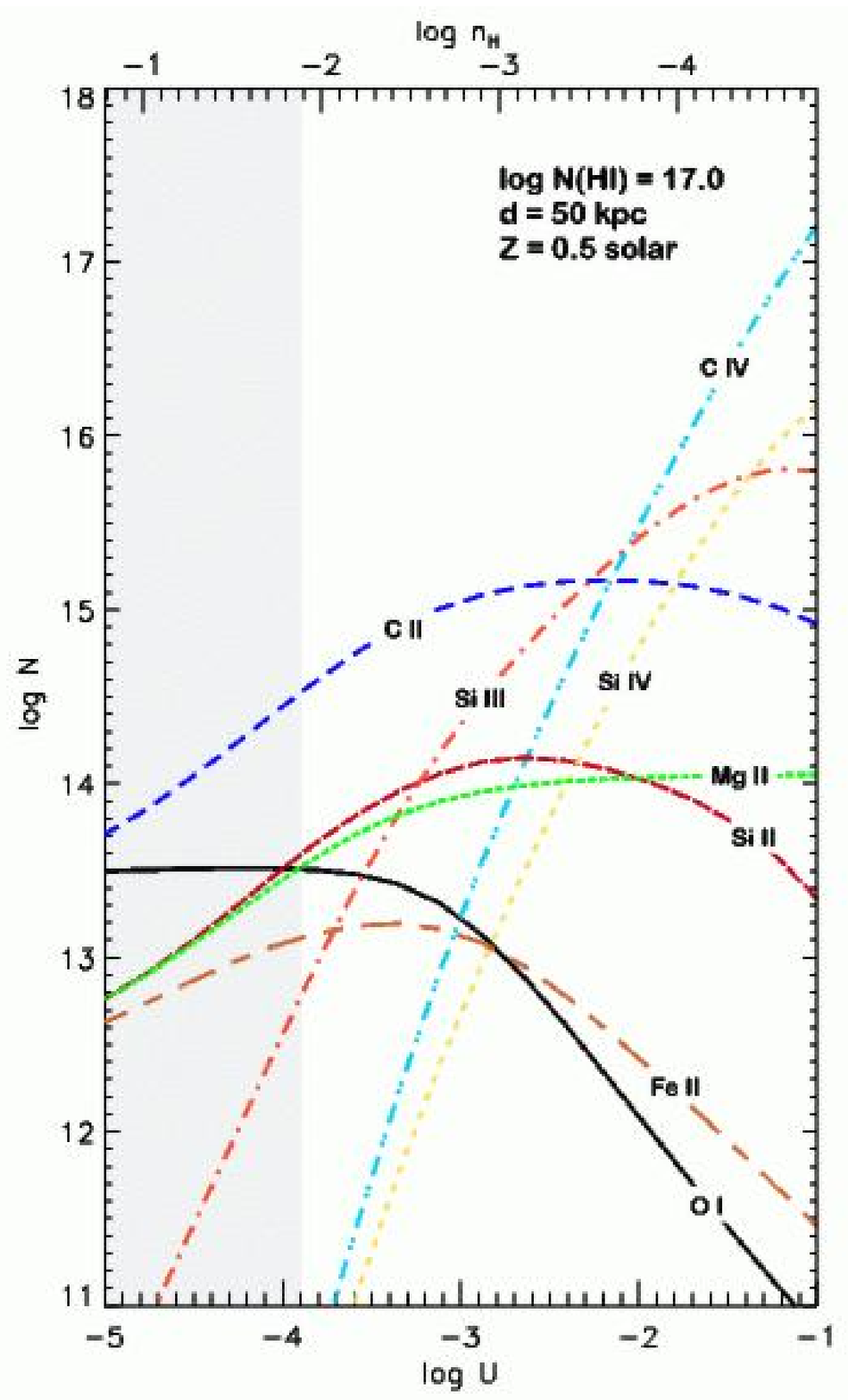}
\caption{An example of a Cloudy photoionization model assuming
log $N$(H\,{\sc i}$)=17.0$, a metallicity of $0.5$ solar, and a distance
of the absorber from the Milky Way disk of $50$ kpc, is shown. The plot 
displays the predicted column densities (log $N$) for various ions as a
function of the hydrogen volume density (log $n_{\rm H}$; upper x axis) and
the ionization parameter (log $U$; lower x axis). The gray-shaded area indicates
the parameter range appropriate for our weak halo absorption systems 
(log [$N$(Si\,{\sc ii})/$N$(O\,{\sc i}$)]\lesssim 0$, which in this model
corresponds to log $n_{\rm H}>-1.8$ and log $U<-3.9$). A HIGH-RESOLUTION
VERSION OF THIS FIGURE IS AVAILABLE ON REQUEST.
}
\end{figure}

\clearpage

\begin{figure}
\epsscale{0.7}
\plotone{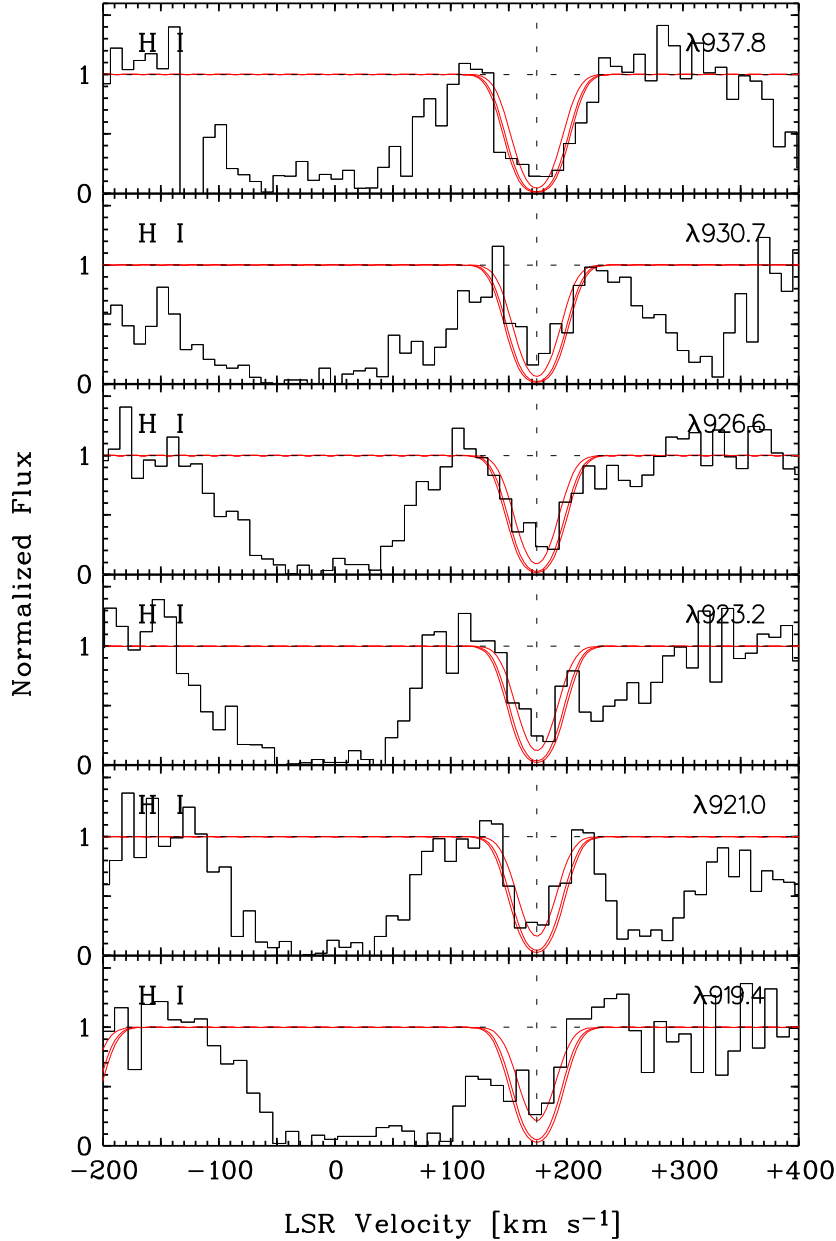}
\caption{
Continuum-normalized absorption profiles of higher order H\,{\sc i} Lyman series lines 
towards PG\,1211+143 are shown, as observed with FUSE. High-velocity
H\,{\sc i} absorption is seen in all lines near $+174$ km\,s$^{-1}$
(indicated with the dashed line). The red solid lines show model spectra
for logarithmic H\,{\sc i} column densities of 16.41, 16.71, and 17.41
(where the strongest absorption corresponds to the highest column densitity).
The best-fitting model is the one with log $N$(H\,{\sc i}$)=16.41$.
}
\end{figure}







\clearpage
\begin{deluxetable}{lccclccl}
\rotate
\tabletypesize{\footnotesize}
\tablewidth{0pt}
\tablecaption{Summary of QSO properties and HST/STIS observations$^{\rm a}$}
\tablehead{

\colhead{}               & 
\colhead{}               &
\colhead{$l$}            &
\colhead{$b$}            & 
\colhead{}               &
\colhead{HVC}  & 
\colhead{O\,{\sc i} $\lambda 1302.2$}     &
\colhead{detected}       \\

\colhead{Name}           & 
\colhead{$z_{\rm em}$}   &
\colhead{(deg)}          &
\colhead{(deg)}          &            
\colhead{Grating}        &
\colhead{absorption$^{\rm b}$}     &
\colhead{strength$^{\rm c}$ (HVC)} &
\colhead{HVC ions$^{\rm d}$}           \\
}

\startdata
PG\,1211+143     & 0.081 & $268$ &  $+74$ & E\,140M          &  yes  &  weak  & O\,{\sc i}, C\,{\sc ii}, Si\,{\sc ii}, Si\,{\sc iii}, C\,{\sc iv} \\
RX\,J1230.8+0115 & 0.117 & $291$ &  $+64$ & E\,140M          &  yes  &  weak  & O\,{\sc i}, C\,{\sc ii}, Si\,{\sc ii}, Si\,{\sc iii}, C\,{\sc iv} \\
NGC\,4151        & 0.003 & $155$ &  $+75$ & E\,140M, E\,230M &  yes  &  weak  & O\,{\sc i}, C\,{\sc ii}, Si\,{\sc ii}, Mg\,{\sc ii}, Fe\,{\sc ii}, Si\,{\sc iii}, C\,{\sc iv}, Si\,{\sc iv} \\
Ton\,S210        & 0.117 & $225$ &  $-83$ & E\,140M, E\,230M &  yes  &  weak  & O\,{\sc i}, C\,{\sc ii}, Si\,{\sc ii}, Mg\,{\sc ii}, Fe\,{\sc ii}, Si\,{\sc iii}, C\,{\sc iv}, Si\,{\sc iv} \\
PHL\,1811        & 0.192 & $47$  &  $-45$ & E\,140M          &  yes  &  weak  & O\,{\sc i}, C\,{\sc ii}, Si\,{\sc ii}, Si\,{\sc iii}, C\,{\sc iv}, Si\,{\sc iv} \\
PG\,1116+215$^{\rm e}$     & 0.177 & $223$ &  $+68$ & E\,140M, E\,230M &  yes &   weak  & O\,{\sc i}, C\,{\sc ii}, Si\,{\sc ii}, Mg\,{\sc ii}, Fe\,{\sc ii}, Si\,{\sc iii}, C\,{\sc iv}, Si\,{\sc iv} \\
\\
\hline
\\
PKS\,0405-123    & 0.574 & $205$ &  $-42$ & E\,140M          &  no  &   null & \nodata \\
HS\,0624+6907    & 0.374 & $146$ &  $+23$ & E\,140M          &  no  &   null & \nodata \\
PG\,0953+415     & 0.239 & $180$ &  $+52$ & E\,140M          &  yes &   null & C\,{\sc ii}, Si\,{\sc iii} \\
Ton\,28          & 0.329 & $200$ &  $+53$ & E\,140M          &  no  &   null & \nodata \\
PG\,1216+069     & 0.334 & $281$ &  $+68$ & E\,140M          &  yes &   null & C\,{\sc ii}, Si\,{\sc ii}, Si\,{\sc iii}, C\,{\sc iv} \\
3C\,273          & 0.158 & $290$ &  $+64$ & E\,140M          &  no  &   null & \nodata \\
PKS\,1302-102    & 0.286 & $309$ &  $+52$ & E\,140M          &  no  &   null & \nodata \\
NGC\,5548        & 0.017 & $32$  &  $+70$ & E\,140M, E\,230M &  no  &   null & \nodata \\
PG\,1444+407     & 0.267 & $70$  &  $+63$ & E\,140M          &  no  &   null & \nodata \\
Mrk\,509         & 0.034 & $36$  &  $-30$ & E\,140M, E\,230M &  yes &   null & C\,{\sc ii}, Si\,{\sc ii}, Si\,{\sc iii}, C\,{\sc iv}, Si\,{\sc iv} \\
PKS\,2155-304    & 0.117 & $18$  &  $-52$ & E\,140M          &  yes &   null & C\,{\sc ii}, Si\,{\sc ii}, Si\,{\sc iii}, C\,{\sc iv}, Si\,{\sc iv} \\
\\
\hline
\\
PKS\,0312-77     & 0.223 & $293$ &  $-38$ & E\,140M, E\,230M &  yes &   strong  & O\,{\sc i}, C\,{\sc ii}, Si\,{\sc ii}, Mg\,{\sc ii}, Fe\,{\sc ii}, Si\,{\sc iii}, C\,{\sc iv}, Si\,{\sc iv} \\ 
3C\,249.1        & 0.313 & $130$ &  $+39$ & E\,140M          &  yes &   blended & C\,{\sc ii}, Si\,{\sc ii}, Si\,{\sc iii} \\
NGC\,3783        & 0.010 & $287$ &  $+23$ & E\,140M, E\,230M &  yes &   strong  & O\,{\sc i}, C\,{\sc ii}, Si\,{\sc ii}, Mg\,{\sc ii}, Fe\,{\sc ii}, Si\,{\sc iii}, C\,{\sc iv}, Si\,{\sc iv} \\
Mrk\,205         & 0.070 & $125$ &  $+42$ & E\,140M          &  yes &   strong  & O\,{\sc i}, C\,{\sc ii}, Si\,{\sc ii}, Si\,{\sc iii} \\
PG\,1259+593     & 0.472 & $121$ &  $+58$ & E\,140M          &  yes &   strong  & O\,{\sc i}, C\,{\sc ii}, Si\,{\sc ii}, Fe\,{\sc ii}, Si\,{\sc iii}, C\,{\sc iv}, Si\,{\sc iv} \\
Mrk\,279         & 0.031 & $115$ &  $+47$ & E\,140M          &  yes &   strong  & O\,{\sc i}, C\,{\sc ii}, Si\,{\sc ii}, Fe\,{\sc ii}, Si\,{\sc iii}, C\,{\sc iv}, Si\,{\sc iv} \\

3C\,351          & 0.372 & $90$  &  $+36$ & E\,140M          &  yes &   strong  & O\,{\sc i}, C\,{\sc ii}, Si\,{\sc ii}, Fe\,{\sc ii}, Si\,{\sc iii}, C\,{\sc iv}, Si\,{\sc iv} \\
H\,1821+643      & 0.297 & $94$  &  $+27$ & E\,140M          &  yes &   strong  & O\,{\sc i}, C\,{\sc ii}, Si\,{\sc ii}, Fe\,{\sc ii}, Si\,{\sc iii}, C\,{\sc iv}, Si\,{\sc iv} \\
NGC\,7469$^{\rm f}$        & 0.016 & $83$  &  $-45$ & E\,140M          &  yes &   weak    & O\,{\sc i}, C\,{\sc ii}, Si\,{\sc ii}, Si\,{\sc iii}, C\,{\sc iv}, Si\,{\sc iv} \\

\enddata
\tablenotetext{a}{The first section of this table lists lines of sight where isolated, high-velocity LLS have been detected in O\,{\sc i} $\lambda 1302.2$
absorption. The second section lists unblocked sightlines that have significant non-detections in high-velocity O\,{\sc i} $\lambda 1302.2$.
The last section lists all other sightlines in our STIS sample; here, strong HVC absorption is present in all cases, thus possibly blending 
absorption from weaker high-velocity LLS along these sightlines.}
\tablenotetext{b}{Only high-velocity absorbers with $|v_{\rm LSR}|>100$ km\,s$^{-1}$ are considered.}
\tablenotetext{c}{O\,{\sc i} $\lambda 1302.2$ line strengths with central absorption depths of (0/$<$1/$\geq$1) are labeled as (null/weak/strong).}
\tablenotetext{d}{HVC detections from ion transitions of oxygen, carbon, silicon, magnesium, and iron are listed.}
\tablenotetext{e}{HVC LLS absorption along this sightline has been analyzed in detail by Ganguly et al.\,(2005).}
\tablenotetext{f}{HVC LLS absorption near $v_{\rm LSR}=-350$ km\,s$^{-1}$ is caused by the ionized envelope of the Magellanic Stream.}
\end{deluxetable}

\clearpage
\begin{deluxetable}{lllllrr}
\tablewidth{0pt}
\tablecaption{Summary of column-density measurements}
\tablehead{
\colhead{}               & \colhead{$l$}    & \colhead{$b$} &
\colhead{$v_{\rm LSR}$}    & \colhead{}       &
\colhead{log $N$}        & \colhead{$b$} \\
\colhead{Name}           & \colhead{(deg)}  & \colhead{(deg)} &
\colhead{(km\,s$^{-1}$)} & \colhead{Ion}    &
\colhead{($N$ in cm$^{-2}$)}                & \colhead{(km\,s$^{-1}$)}
}
\startdata
PG\,1211+143        & $268$ &  $+74$ & $+174$ & C\,{\sc ii}   & $13.63\pm 0.06^{\rm a}$ & $7.5 \pm 2.1$ \\
                    &       &        &        & O\,{\sc i}    & $13.10\pm 0.03$ & $7.5 \pm 2.1$ \\
                    &       &        &        & Si\,{\sc ii}  & $12.63\pm 0.03$ & $7.5 \pm 2.1$ \\
                    &       &        &        & Fe\,{\sc ii}  & $<13.03$        & $7.5$ \\
                    &       &        &        & Si\,{\sc iii} & $12.19\pm 0.036^{\rm a}$ & $7.5 \pm 2.1$ \\
	            &       &        &        & C\,{\sc iv}   & $12.46\pm 0.05$ & $7.5 \pm 2.1$ \\
                    &       &        &        & Si\,{\sc iv}  & $<12.60$        & $7.5$ \\
\\
	            &       &        & $+191$ & C\,{\sc ii}   & $13.00\pm 0.04$ & $6.9 \pm 3.3$ \\
                    &       &        &        & Si\,{\sc iii} & $12.12\pm 0.03$ & $6.9 \pm 3.3$ \\
	            &       &        &        & C\,{\sc iv}   & $12.67\pm 0.05$ & $6.9 \pm 3.3$ \\
\\
RX\,J1230.8+0115    & $291$ &  $+64$ & $+293$ & C\,{\sc ii}   & $15.10\pm 0.16$ & $4.0 \pm 1.7$ \\
                    &       &        &        & O\,{\sc i}    & $13.80\pm 0.03$ & $4.0 \pm 1.7$ \\
	            &       &        &        & Si\,{\sc ii}  & $13.30\pm 0.03$ & $4.0 \pm 1.7$ \\
                    &       &        &        & Fe\,{\sc ii}  & $<13.24$        & $4.0$ \\
	            &       &        &        & Si\,{\sc iii} & $12.59\pm 0.06$ & $10.0 \pm 3.0$ \\
                    &       &        &        & C\,{\sc iv}   & $12.99\pm 0.07$ & $10.0 \pm 3.0$ \\
	            &       &        &        & Si\,{\sc iv}  & $<11.61$        & $10.0$ \\
\\
                    &       &        & $+110$ & C\,{\sc ii}   & $13.75\pm 0.05$ & $8.0 \pm 2.8$ \\
	            &       &        &        & O\,{\sc i}    & $<13.65$        & $8.0$ \\
                    &       &        &        & Si\,{\sc ii}  & $12.68\pm 0.07$ & $8.0 \pm 2.8$ \\
                    &       &        &        & Si\,{\sc iii} & $12.62\pm 0.06$ & $8.0 \pm 2.8$ \\
\\
NGC\,4151           & $155$ &  $+75$ & $+145$ & C\,{\sc ii}   & $13.78\pm 0.05$ & $3.0 \pm 1.4$ \\
                    &       &        &        & O\,{\sc i}    & $12.90\pm 0.03$ & $3.0 \pm 1.4$ \\
                    &       &        &        & Si\,{\sc ii}  & $12.72\pm 0.03$ & $3.0 \pm 1.4$ \\
	            &       &        &        & Mg\,{\sc ii}  & $12.19\pm 0.04$ & $3.0 \pm 1.4$ \\
                    &       &        &        & Fe\,{\sc ii}  & $12.49\pm 0.05$ & $3.0 \pm 1.4$ \\
	            &       &        &        & Si\,{\sc iii} & $12.70\pm 0.05$ & $11.0 \pm 2.8$ \\
                    &       &        &        & C\,{\sc iv}   & $13.56\pm 0.05$ & $16.3 \pm 4.4$ \\
	            &       &        &        & Si\,{\sc iv}  & $12.73\pm 0.06$ & $16.3 \pm 4.4$ \\
\\
Ton\,S210           & $225$ &  $-83$ & $-167$ & C\,{\sc ii}   & $15.20\pm 0.256^{a}$ & $6.2 \pm 2.3$ \\
                    &       &        &        & O\,{\sc i}    & $14.02\pm 0.04$ & $6.2 \pm 2.3$ \\
                    &       &        &        & Si\,{\sc ii}  & $13.64\pm 0.04$ & $6.2 \pm 2.3$ \\
	            &       &        &        & Mg\,{\sc ii}  & $12.95\pm 0.07$ & $6.2 \pm 2.3$ \\
	            &       &        &        & Fe\,{\sc ii}  & $13.95\pm 0.06$ & $6.2 \pm 2.3$ \\
	            &       &        &        & Si\,{\sc iii} & \nodata $^{\rm b}$  & \nodata \\
	            &       &        &        & C\,{\sc iv}   & $<12.93$        & $6.2$ \\
	            &       &        &        & Si\,{\sc iv}  & $<12.13$        & $6.2$ \\
\\
                    &       &        & $-188$ & C\,{\sc ii}   & $14.00\pm 0.11$ & $12.3 \pm 4.6$ \\
\\
                    &       &        & $-210$ & C\,{\sc ii}   & $13.30\pm 0.07$ & $6.0 \pm 1.8$ \\
\\
                    &       &        & $-242$ & C\,{\sc ii}   & $12.97\pm 0.04$ & $4.0 \pm 1.2$ \\
                    &       &        &        & Si\,{\sc iii} & $12.83\pm 0.05$ & $7.6 \pm 2.0$ \\
	            &       &        &        & C\,{\sc iv}   & $13.78\pm 0.06$ & $7.6 \pm 2.0$ \\
                    &       &        &        & Si\,{\sc iv}  & $12.76\pm 0.04$ & $7.6 \pm 2.0$ \\
\\
PHL\,1811           &  $47$ &  $-45$ & $-207$ & C\,{\sc ii}   & $14.70\pm 0.21$ & $9.2 \pm 2.9$ \\
	            &       &        &        & O\,{\sc i}    & $14.11\pm 0.05$ & $9.2 \pm 2.9$ \\
                    &       &        &        & Si\,{\sc ii}  & $13.74\pm 0.04$ & $9.2 \pm 2.9$ \\
                    &       &        &        & Al\,{\sc ii}  & $12.60\pm 0.06$ & $9.2 \pm 2.9$ \\
	            &       &        &        & Fe\,{\sc ii}  & $<12.49$        & $9.2$ \\
	            &       &        &        & Si\,{\sc iii} & \nodata $^{\rm b}$  & \nodata \\
	            &       &        &        & C\,{\sc iv}   & $13.27\pm 0.12$ & $9.2 \pm 2.9$ \\
	            &       &        &        & Si\,{\sc iv}  & $12.77\pm 0.11$ & $9.2 \pm 2.9$ \\
\\
                    &       &        & $-163$ & C\,{\sc ii}   & $13.92\pm 0.09$ & $9.6 \pm 3.2$ \\
	            &       &        &        & O\,{\sc i}    & $13.23\pm 0.07$ & $9.6 \pm 3.2$ \\
	            &       &        &        & Si\,{\sc ii}  & $12.58\pm 0.05$ & $9.6 \pm 3.2$ \\
	            &       &        &        & Al\,{\sc ii}  & $12.16\pm 0.09$ & $9.6 \pm 3.2$ \\
	            &       &        &        & Fe\,{\sc ii}  & $<13.18$        & $9.6$ \\
	            &       &        &        & Si\,{\sc iii} & \nodata $^{\rm b}$  & \nodata \\
	            &       &        &        & C\,{\sc iv}   & \nodata $^{\rm b}$  & \nodata \\
	            &       &        &        & Si\,{\sc iv}  & $13.51\pm 0.11$ & $10.6 \pm 4.1$ \\
\\
                    &       &        & $-264$ & C\,{\sc ii}   & $13.65\pm 0.13^{a}$ & $19.9 \pm 6.3$ \\
	            &       &        &        & Si\,{\sc iii} & \nodata $^{\rm b}$  & \nodata \\
\\
	            &       &        & $-350$ & Si\,{\sc iii} & \nodata $^{\rm b}$  & \nodata \\
                    &       &        &        & C\,{\sc iv}   & $13.83\pm 0.09$ & $14.2 \pm 3.6$ \\
	            &       &        &        & Si\,{\sc iv}  & $12.69\pm 0.05$ & $14.2 \pm 3.6$ \\
\enddata
																			    \tablenotetext{a}{Absorption profile asymmetric; evidence for unresolved velocity-component strucure.}\\
																			    \tablenotetext{b}{Absorption fully saturated; reliable column-density estimate not possible.}
																			    \end{deluxetable}

\clearpage
\begin{deluxetable}{lccc}
\tablewidth{0pt}
\tablecaption{Parameters of photoionization modeling$^{\rm a}$}
\tablehead{
\colhead{}                    & \colhead{$v_{\rm LSR}$}  &
\colhead{log $N$(H\,{\sc i})} &  \colhead{} \\
\colhead{Direction}           & \colhead{(km\,s$^{-1}$)} &
\colhead{($N$ in cm$^{-2})$}  & \colhead{log $\frac{N({\rm Si\,II})}{N({\rm O\,I})}$}  \\
}
\startdata
PG\,1211+143     & $+174$ & $16.41\,...\,17.41$ & $-0.47$ \\
RX\,J1230.8+0115 & $+293$ & $17.11\,...\,18.11$ & $-0.50$ \\
NGC\,4151        & $+145$ & $16.21\,...\,17.21$ & $-0.18$ \\ 
Ton\,S210        & $-167$ & $17.33\,...\,18.33$ & $-0.38$ \\
PHL\,1811        & $-207$ & $17.42\,...\,18.42$ & $-0.37$ \\
                 & $-163$ & $16.54\,...\,17.54$ & $-0.65$ \\
PG\,1116+215     & $+193$ & $17.13\,...\,18.13$ & $+0.03$ \\		 
\enddata
\tablenotetext{a}{Assuming a metallicity of the gas of $0.1-1.0$ solar and a silicon depletion
into dust grains of $A_{\rm Si}=0.0-0.4$ dex.}\\
\tablenotetext{b}{$f_{\rm H}=N$(H\,{\sc i}+H\,{\sc ii})/$N$(H\,{\sc i}).}
\end{deluxetable}

\clearpage
\begin{deluxetable}{lcccccc}
\tabletypesize{\small}
\tablewidth{0pt}
\tablecaption{Physical parameters for cloud model I$^{\rm a}$}
\tablehead{
\colhead{}                    & \colhead{$v_{\rm LSR}$}     &
\colhead{log $N$(H\,{\sc i})} & \colhead{}                &
\colhead{}                    & \colhead{log $n_{\rm H}$}  &
\colhead{log $L$}             \\
\colhead{Direction}           & \colhead{(km\,s$^{-1}$)}  &
\colhead{($N$ in cm$^{-2})$}  & \colhead{log $U$}         &
\colhead{log $f_{\rm H}$$^{\rm b}$}     & \colhead{($n_{\rm H}$ in cm$^{-3})$}  &
\colhead{($L$ in pc)}         \\
}
\startdata
PG\,1211+143     & $+174$ & $16.71$ & $-4.83$ & $0.65$ & $-0.83$ & $-0.31$ \\
RX\,J1230.8+0115 & $+293$ & $17.41$ & $-4.81$ & $0.67$ & $-0.85$ & $-0.45$ \\
NGC\,4151        & $+145$ & $16.51$ & $-4.35$ & $1.07$ & $-1.31$ & $+0.40$ \\
Ton\,S210        & $-167$ & $17.63$ & $-4.45$ & $0.98$ & $-1.21$ & $+1.34$ \\
PHL\,1811        & $-207$ & $17.72$ & $-4.37$ & $1.06$ & $-1.29$ & $+1.58$ \\
                 & $-163$ & $16.84$ & $-5.14$ & $0.38$ & $-0.52$ & $-0.74$ \\
PG\,1116+215     & $+193$ & $17.43$ & $-3.91$ & $1.46$ & $-1.71$ & $+2.11$ \\
\enddata
\tablenotetext{a}{Model I assumes a metallicity of $0.5$ solar, a cloud distance
from the Galaxy of $d=50$ kpc and no depletion of Si onto dust grains ($A_{\rm Si}=0$).}\\
\tablenotetext{b}{$f_{\rm H}=N$(H\,{\sc i}+H\,{\sc ii})/$N$(H\,{\sc i}).}
\end{deluxetable}

\clearpage
\begin{deluxetable}{lccc}
\tabletypesize{\small}
\tablewidth{0pt}
\tablecaption{Selected physical parameters for cloud model II$^{\rm a}$}
\tablehead{
\colhead{}                    & \colhead{$v_{\rm LSR}$}     &
\colhead{log $n_{\rm H}$}  &
\colhead{log $L$}             \\
\colhead{Direction}           & \colhead{(km\,s$^{-1}$)}  &
\colhead{($n_{\rm H}$ in cm$^{-3})$}  &
\colhead{($L$ in pc)}         \\
}
\startdata
PG\,1211+143     & $+174$ & $-0.03$ & $-1.11$ \\
RX\,J1230.8+0115 & $+293$ & $-0.05$ & $-0.35$ \\
NGC\,4151        & $+145$ & $-0.51$ & $-0.40$ \\
Ton\,S210        & $-167$ & $-0.41$ & $+0.54$ \\
PHL\,1811        & $-207$ & $-0.49$ & $+0.78$ \\
                 & $-163$ & $+0.28$ & $-1.54$ \\
PG\,1116+215     & $+193$ & $-0.95$ & $+1.35$ \\
\enddata
\tablenotetext{a}{Model II assumes a metallicity of $0.5$ solar, a cloud distance
from the Galaxy of $d=10$ kpc and no depletion of Si onto dust grains ($A_{\rm Si}=0$).
Note that the values for $N$(H\,{\sc i}), log $U$, and log $f_{\rm H}$ are the same
as for cloud model I (see Table 2).} 
\end{deluxetable}

\end{document}